\documentclass[aps, prd, twocolumn, preprintnumbers, groupedaddress, nofootinbib, amssymb, notitlepage, eqsecnum]{revtex4-2}

\usepackage{bm}
\usepackage[dvipsnames]{xcolor}
\usepackage{amsmath,amsthm,amssymb}
\usepackage{amsfonts}
\usepackage{hyperref}
\usepackage{graphicx}

\allowdisplaybreaks[1]

\newcommand{\rd}{{\rm d}}
\newcommand{\be}{\begin{equation}}
\newcommand{\ee}{\end{equation}}
\newcommand{\ba}{\begin{eqnarray}}
\newcommand{\ea}{\end{eqnarray}}
\newcommand{\Mpl}{M_{\rm Pl}}

\begin{document}

\preprint{YITP-26-20, WUCG-26-02}

\title{Light rings, gravitational lensing, 
and ISCOs 
of exotic compact objects \\
in Einstein-scalar-Maxwell theories}

\author{Antonio De Felice$^{a}$}
\email{antonio.defelice@yukawa.kyoto-u.ac.jp}  
\author{Shinji Tsujikawa$^{b}$}
\email{tsujikawa@waseda.jp} 

\affiliation{
$^{a}$Center for Gravitational Physics and Quantum Information, Yukawa Institute for Theoretical Physics, Kyoto University, 606-8502, Kyoto, Japan\\
$^{b}$Department of Physics, Waseda University, 
3-4-1 Okubo, Shinjuku, Tokyo 169-8555, Japan}
 
\date{\today}

\begin{abstract}

In Einstein-scalar-Maxwell theories with a coupling between the scalar field $\phi$ and the electromagnetic field strength $F$ of the form $\mu(\phi) F$, we investigate the existence of 
exotic compact objects (ECOs) and their observational signatures in photon and massive-particle dynamics. For $\mu(\phi)$ diverging at the origin while all physical quantities remain finite, we demonstrate the existence of electrically charged ECOs with a shell-like structure whose density peaks at an intermediate radius. We compute their mass and radius, together with the scalar and vector field profiles, on a static and spherically symmetric background. We then examine the existence of light rings and place bounds on a model parameter by requiring the absence of a linearly stable light ring. Under this condition, photon echoes from ECOs are absent. We also compute the gravitational-lensing deflection angle and show that it attains a maximum for an impact parameter of the same order as the ECO radius.
Finally, we study the parameter space in which innermost stable circular orbits of massive particles exist.

\end{abstract}


\maketitle

\section{Introduction}
\label{Intro}

General Relativity (GR) has withstood more than a century of experimental scrutiny, ranging from Solar System tests \cite{Will1993,Will:2014kxa} to direct detections of gravitational waves \cite{LIGOScientific:2016aoc,LIGOScientific:2017vwq}.
However, when applied to phenomena on galactic and cosmological scales, GR appears to require the existence of a dominant, 
non-luminous component---dark matter (DM) \cite{Blumenthal:1984bp,Jungman:1995df,Bertone:2004pz}.
The empirical evidence for DM is inferred from galaxy rotation curves \cite{Zwicky:1937zza,Rubin:1970zza,Sofue:2000jx}, gravitational lensing observations \cite{Bartelmann:1999yn,Clowe:2006eq}, the formation and evolution of large-scale structure \cite{Anderson2014,eBOSS2021,DES:2024jxu,DESI:2025zgx}, 
and measurements of the cosmic microwave background \cite{WMAP:2012nax,Planck:2018vyg}.
Collectively, these observations consistently indicate that approximately $85\,\%$ of the non-relativistic matter content of the Universe is non-baryonic and effectively collisionless.
Despite its overwhelming gravitational influence, the microscopic nature of DM remains unknown.

DM may originate from scalar or vector fields arising in theories beyond the Standard Model of particle physics. 
For instance, string theory predicts the existence of a dilaton field $\phi$ \cite{Yoneya:1974jg,Scherk:1974ca,Callan:1986bc}, 
which can couple to the electromagnetic field strength 
$F$ through the term $e^{-\lambda \phi}F$. 
This scalar-vector interaction can be extended to a more general coupling of the form $\mu(\phi)F$, 
where $\mu$ is an arbitrary function of  
the scalar field $\phi$.
The simplest framework for massless fields, which interact with each other but are minimally coupled to gravity, is provided by Einstein-scalar-Maxwell (ESM) theories with the Lagrangian $\Mpl^2 R/2+X+\mu(\phi)F$ \cite{Gaillard:1981rj,Duff:1995sm,Andrianopoli:1996cm}, 
where $\Mpl$ denotes the reduced Planck mass, 
$R$ is the Ricci scalar, and
$X=-\partial_{\mu}\phi \partial^{\mu}\phi/2$ 
represents the kinetic term of the scalar field.
In local regions of the 
Universe, scalar and vector fields may undergo gravitational clustering, leading to the formation of hairy black holes (BHs) or exotic 
compact objects (ECOs). 

In Einstein-dilaton-Maxwell theories with the exponential coupling $\mu(\phi)=\mu_0 e^{-\lambda \phi}$, there exists an exact, static, spherically symmetric BH solution carrying both scalar and electric (or magnetic) charges \cite{Gibbons:1987ps,Garfinkle:1990qj}. This solution possesses a curvature singularity at the center of the BH. 
In Ref.~\cite{DeFelice:2024ops}, the authors attempted to obtain regular black hole (BH) solutions within a general class of scalar-vector-tensor theories described by the Lagrangian $\Mpl^2 R/2+{\cal L}(\phi,X,F)$. 
By requiring that the BH solutions be stable against linear perturbations near the center, they did not find any example of a regular BH solution free from instabilities, including the case of nonlinear electrodynamics \cite{DeFelice:2024seu}.

If the existence of horizons is not required, it is in principle possible to construct ECOs with regular centers through the gravitational clustering of scalar or vector fields (see Refs.~\cite{Schunck:2003kk,Liebling:2012fv,Barack:2018yly,Cardoso:2019rvt} for reviews). 
Nevertheless, the construction of stable, regular compact objects based on {\it real} fields is, in general, a nontrivial problem. In GR with a real scalar field described by the Lagrangian $\Mpl^2 R/2 + {\cal L}(\phi, X)$, it is known that no static, spherically symmetric, regular solitonic solutions with positive-definite energy density exist in the absence of ghosts \cite{Derrick:1964ww,Diez-Tejedor:2013sza}.
The nonexistence of stable compact objects likewise applies to a real Abelian vector field in electromagnetism \cite{Wheeler:1955zz,Power:1957zz}. 
By contrast, when complex scalar or vector fields endowed with mass terms are considered, self-gravitating configurations such as boson stars \cite{Kaup:1968zz,Ruffini:1969qy} or Proca stars \cite{Brito:2015pxa,Herdeiro:2016tmi} can exist. 
These results suggest that the realization of regular compact objects based on real fields requires the interplay between scalar and vector degrees of freedom, as in ESM theories.

If the coupling $\mu(\phi)$ in ESM theories remains finite at all radii $r$, it has been shown in Ref.~\cite{Herdeiro:2019oqp} that no asymptotically flat, localized solitonic solutions exist.
If, on the other hand, one allows $\mu(\phi)$ to diverge at $r=0$, it becomes possible to construct electrically charged ECOs
with regular centers \cite{Herdeiro:2019iwl,DeFelice:2024ops,DeFelice:2025vef}.
In this case, although $\mu(\phi)$ diverges at $r=0$, all physical quantities remain finite, and therefore no inconsistency arises. Moreover, the divergence of $\mu(\phi)$ can be regarded as the weak-coupling limit of the theory.
In the magnetically charged case, a similar star configuration can be realized when $\mu(\phi)$ approaches zero as $r \to 0$. In this latter case, however, the ECOs suffer from a strong-coupling problem for perturbations near 
the center \cite{DeFelice:2025vef}.

In Ref.~\cite{DeFelice:2025vef}, the authors assumed analytic forms for the two metric functions, $f(r)$ and $h(r)$, consistent with regularity at $r=0$, and reconstructed the form of $\mu(\phi)$ that realizes these solutions. 
From a theoretical standpoint, it is essential to understand the structural robustness of such hairy configurations: do they persist for a given coupling function $\mu(\phi)$? 
In this paper, we provide the explicit form of $\mu(\phi)$ from the beginning and obtain the density profile of hairy ECOs endowed with both scalar and electric charges. 
For consistency with regular boundary conditions at $r=0$ 
and asymptotic flatness, the coupling is required 
to take the form
$\mu(\phi)=\mu_0+\mu_1 \Mpl^p (\phi-\phi_0)^{-p}$, 
where $\mu_0$ and $\mu_1$ are constants, 
$\phi_0$ is the field value at $r=0$, and the power 
$p$ lies in the range $2<p \le 3$.
As expected, $\mu(\phi)$ diverges as $r \to 0$, 
a behavior that is consistent with 
the findings 
of Ref.~\cite{Herdeiro:2019iwl}.

By reflecting the property that the divergence of $\mu(\phi)$ at $r=0$ corresponds to the weak-coupling limit, the density and pressure of ECOs vanish as $r \to 0$. 
As a result, the ECO exhibits a peculiar shell-like structure, with the density reaching a maximum at an intermediate radius. 
This analysis clarifies the degree of universality in the existence of solitonic compact objects within ESM theories. 
We define the radius of the object based on the thickness around this peak density to obtain the mass-radius relation. 
The resulting compactness of ECOs can be as large as 
${\cal C} \sim 0.01$--$0.1$. These properties are crucial for assessing their distinguishability from BHs or other ECOs.

The distinctive density profile of ECOs in ESM theories intrinsically arises from interactions between scalar and vector fields in the dark sector.
Although these fields do not directly interact with Standard Model particles, their coupling to gravity alters the background geometry in a nontrivial way. 
This affects the trajectories of photons and massive particles in the vicinity of ECOs, which can be distinguished from those around BHs and other ECOs.

To understand the key geometric and astrophysical properties of ECOs in ESM theories, we first analyze the existence of light rings and determine the associated null geodesic structure. Light rings provide a powerful diagnostic of strong-gravity 
compactness \cite{Gralla:2019xty,Gralla:2020srx,Gan:2021xdl,Paugnat:2022qzy,Staelens:2023jgr,Olmo:2023lil}
and are directly relevant to high-resolution observations, such as those performed by the Event 
Horizon Telescope \cite{EventHorizonTelescope:2019dse,Broderick:2022tfu,Broderick:2021ohx}. Theoretically, there can be cases 
in which two light rings appear---one linearly stable and the other unstable. The existence of a linearly stable ring, however, may be subject to nonlinear instability due to the accumulation of photon energy \cite{Cunha:2017qtt,Cunha:2022gde}. 
We show that the model parameters can be constrained by requiring the absence of a linearly stable light ring.

We further investigate the gravitational lensing signatures of ECOs by computing the photon deflection angle $\Psi$ as a function of the impact parameter $b$ (see Refs.~\cite{Bozza:2002zj,Keeton:2005jd,Bozza:2010xqn,Cunha:2018acu,Shaikh:2019itn,AbhishekChowdhuri:2023ekr} for studies of gravitational lensing by compact objects). 
Notably, $\Psi$ attains a maximum for $b$ of the order of the stellar radius and decreases as $r \to 0$ and $r \to \infty$.
We also clarify the parameter space in which the innermost stable circular orbit (ISCO) \cite{Carter:1968rr,Chandrasekhar:1984siy} 
of massive particles exists. Indeed, ISCOs can exist even for parameter values that are constrained by the absence of a linearly stable light ring. Our analysis of light rings, gravitational lensing, and ISCOs may provide avenues to discriminate such ECOs from other compact objects. 

Although the model provides a well-defined theoretical framework and exhibits rich phenomenology at the level of geodesic motion, its full observational and cosmological implications remain largely unexplored. In particular, its stability properties, formation channels, abundance, and potential role as a dominant 
DM component require further investigation.
The present analysis therefore represents a step toward a systematic characterization of this class of dark ECOs, helping to bridge the gap between theoretical construction and phenomenological assessment.

This paper is organized as follows. In Sec.~\ref{backsec}, we present the theoretical framework and introduce an explicit coupling function $\mu(\phi)$ responsible for the construction of regular ECOs. In Sec.~\ref{configsec}, we study the density profiles of ECOs, along with the solutions of the scalar and vector fields, and compute the mass–radius relation. In Secs.~\ref{ringsec}, \ref{lensingsec}, and \ref{ISCOsec}, we analyze light rings, gravitational lensing, and ISCOs, respectively. We conclude in Sec.~\ref{consec} with a discussion of open problems and future directions. In the Appendix, we present the computation of the gravitational lensing deflection angle for the model with analytic metric functions proposed in Ref.~\cite{DeFelice:2024ops}.

\section{Coupling in ESM theories}
\label{backsec}

We consider a real scalar field $\phi$ with 
the kinetic term $X 
=-\partial_{\mu}\phi \partial^{\mu}\phi/2$ 
and a $U(1)$ gauge field $A_{\mu}$ 
characterized with field strength 
$F = -F_{\mu\nu}F^{\mu\nu}/4$, 
where $F_{\mu\nu} = \partial_{\mu}A_{\nu} - \partial_{\nu}A_{\mu}$. 
We assume that these scalar and vector fields 
are present in the dark sector of the Universe. 
The gravitational sector is described by 
GR, whose dynamics is governed by the 
Einstein-Hilbert Lagrangian $\Mpl^{2}R/2$.
The total action is given by 
\be
{\cal S}=\int \rd^4 x \sqrt{-g} 
\left[ \frac{\Mpl^2}{2}R
+X+\mu (\phi)F \right]\,,
\label{action}
\ee
where $g$ denotes the determinant of the metric tensor 
$g_{\mu\nu}$, and $\mu(\phi)$ is the 
scalar-vector coupling that depends on $\phi$.

Let us consider the static and spherically 
symmetric background described 
by the line element 
\be
{\rm d}s^2=-f(r){\rm d}t^2+h^{-1}(r){\rm d}r^2+
r^2 \left({\rm d} \theta^2+\sin^2 \theta\, 
{\rm d}\varphi^2 \right)\,,
\label{line}
\ee
where $f$ and $h$ are functions of $r$. 
We take the scalar and vector field configurations 
in the forms $\phi=\phi(r)$ and $A_{\mu} 
{\rm d}x^{\mu}=A_0(r){\rm d}t$, respectively. 
We are interested in constructing 
regular compact objects with an electric charge $q_E$, 
while setting the magnetic charge to zero ($q_M = 0$).
The quantities $X$ and $F$ 
can be expressed as $X=-(1/2)h \phi'^2$ and 
$F=h A_0'^2/(2f)$, where a prime denotes the 
derivative with respect to $r$.

The (00), (11), and (22) [=(33)] components of 
the gravitational equations of motion are given by
\ba
& &
\frac{\Mpl^2 (rh'+h-1)}{r^2}
=-\rho\,,\label{back1} \\
& &
\frac{\Mpl^2 [rhf'+f(h-1)]}{r^2 f}=P_r\,,
\label{graeq2} \\
& &
\frac{\Mpl^2 [2rfh f'' 
-rh f'^2+ ( rh' + 2h) ff' 
+ 2f^2h']}{4rf^2} 
\nonumber \\
& &
=P_{t}\,, 
\label{graeq3}
\ea
where 
\ba
\rho &\equiv& \frac{1}{2}h \phi'^2
+\frac{h}{2f} \mu A_0'^2\,,\label{rho} \\
P_r &\equiv& \frac{1}{2}h \phi'^2
-\frac{h}{2f} \mu A_0'^2\,,\label{Pr} \\
P_{t} &\equiv& -P_{r}\,.\label{Pt}
\ea
Here, $\rho$ denotes the energy density, 
whereas $P_r$ and $P_t$ 
correspond to the radial and transverse pressures, respectively. 

We introduce the following function: 
\be
N(r)=\frac{f(r)}{h(r)}\,,
\ee
which is required to be positive to ensure that 
$-g>0$. By using Eqs.~(\ref{back1}) and 
(\ref{graeq2}), it follows that 
\be
\phi'=\Mpl \sqrt{\frac{N'}{rN}}\,,
\label{rN}
\ee
where we have chosen the branch $\phi'>0$ 
without loss of generality.
Varying the action (\ref{action}) 
with respect to $A_0$ 
and $\phi$, respectively, we obtain
\ba
& &
A_0'=\frac{q_E \sqrt{N}}{\mu\,r^2}\,,
\label{back3}\\
& &
\left( \sqrt{N} h\,r^2 \phi' \right)'
+\sqrt{N}r^2 \mu_{,\phi} 
\frac{h A_0'^2}{2f}=0\,,
\label{back4}
\ea
where the constant $q_E$ corresponds to the electric charge, 
and $\mu_{,\phi} \equiv {\rm d}\mu/{\rm d}\phi$.
By using Eq.~(\ref{back1}) with Eqs.~(\ref{rho}), (\ref{rN}), 
and (\ref{back3}), the coupling $\mu$ can be expressed as 
\be
\mu=-\frac{q_E^2 N}
{\Mpl^2 r^2 [2(rh'+h-1)N+rhN']}\,.
\label{mu2}
\ee
We substitute Eqs.~(\ref{rN}) and (\ref{back3}) into Eq.~(\ref{back4}) to eliminate the terms $\phi'$ and $A_0'$. 
We also take the derivative of Eq.~(\ref{mu2}) with respect to $r$ and substitute the resulting expression 
for $\mu_{,\phi}$ into Eq.~(\ref{back4}). 
We thus obtain the following differential equation:
\ba
& &
2 \left( r^2 h''+4r h'+2h-2 
\right)N^2-r^2 h N'^2 
\nonumber \\
& &
+3r( rh'+2h) NN'
+2r^2 h N N''=0\,.
\label{hNeq}
\ea
To ensure the existence of compact objects that 
are regular at the center, the metric functions 
must be expanded as
\ba
h(r) &=& 1+\sum_{n=2}^{\infty} h_n r^n\,,
\label{hexpan} \\
N(r) &=& N_0+\sum_{n=2}^{\infty} N_n r^n\,,
\ea
where $h_n$, $N_0$, and $N_n$ are constants.

The solution to Eq.~(\ref{hNeq}) that is 
consistent with the regularity 
condition (\ref{hexpan}) is given by 
\be
h(r)=\frac{2 \int_0^r \sqrt{N(r_1)} 
(\int_0^{r_1} \sqrt{N(r_2)} 
{\rm d}r_2) {\rm d}r_1}
{r^2 N(r)}\,, 
\label{hr2}
\ee
where the two integration constants have been 
set to zero to ensure regularity at $r = 0$.
Since $N(r) > 0$, $h(r)$ remains positive 
for all $r > 0$, implying the absence of horizons. 
Therefore, in the following, we focus on 
horizonless regular compact objects.
 
By considering the function $N(r)=N_0+N_n r^n$ 
near $r=0$, where $N_0$, $N_n$, and $n$ are 
constants, the leading-order scalar derivative 
is given by 
$\phi'=\Mpl \sqrt{n N_n/N_0}\,r^{(n-2)/2}$.
To prevent the formation of a cusp-like structure, 
we impose the boundary condition $\phi'(r=0)=0$, which requires $n>2$. Furthermore, if $n$ is an odd integer, 
$\phi'$ becomes non-analytic; for instance, 
for $n=3$, one finds $\phi' \propto r^{1/2}$.
For consistency, we therefore restrict $n$ 
to be an even integer with $n \geq 4$. 
In this case, the expansion of $N(r)$ 
near $r=0$ reads
\be
N(r)= N_0+\sum_{m=2}^{\infty} N_{2m} r^{2m}\,,
\label{Nexpan}
\ee
where the sum runs over integers $m \ge 2$. 
For $N(r) = N_0 + N_{2m} r^{2m}$, 
the scalar field near $r = 0$ behaves as 
\be
\phi(r) \simeq 
\phi_0+\Mpl \sqrt{\frac{2 N_{2m}}{m N_0}}r^m\,,
\label{phia}
\ee
where $\phi_0$ is an integration constant 
corresponding to the field value at $r=0$.
Substituting $N(r)=N_0+N_{2m} r^{2m}$ 
into Eq.~(\ref{hr2}) and integrating it near $r=0$, 
the metric function $h$ behaves as 
\be
h(r) \simeq 1-\frac{2mN_{2m}}{(2m+1)N_0}r^{2m}\,.
\label{ha}
\ee
From Eq.~(\ref{mu2}), the leading-order term of $\mu$ 
around $r=0$ is given by 
\be
\mu (r) \simeq \frac{q_E^2 N_0}{2\Mpl^2 m N_{2m}} 
r^{-2(m+1)}\,.
\label{mua}
\ee
From Eqs.~(\ref{phia}) and (\ref{mua}), 
the coupling behaves near $r=0$ as
\be
\mu(\phi) \propto \left( \phi-\phi_0 
\right)^{-2(m+1)/m}\,.
\label{mur=0}
\ee

An example of $N(r)$ whose expansion around $r=0$ 
starts at order $m=2$ was proposed in Refs.~\cite{DeFelice:2024ops,DeFelice:2025vef}. 
It takes the form
$N(r)=(r^4+\sqrt{N_0}r_0^4)^2/(r^4+r_0^4)^2$.
Near $r=0$, this function admits the expansion  
$N(r)=N_0+2(\sqrt{N}_0-N_0)r^4/r_0^4+{\cal O}(r^8)$. 
At spatial infinity, $N(r)\to 1$, thereby recovering the asymptotically flat solution with $f=h\to 1$. 
Expanding this form of $N(r)$ at large distances, we find that the coupling $\mu(\phi)$ approaches a constant 
value $\mu_{\infty}$ in the limit $r \to \infty$. 
More precisely, it behaves as 
$\mu(\phi) \simeq \mu_{\infty}[1+\beta (\phi-\phi_{\infty})]$, 
where $\beta$ is a constant and $\phi_{\infty}$ 
denotes the field value at spatial infinity. 
Note that $\phi_{\infty}$ determines the vacuum expectation value (VEV) of the field at spatial infinity. 

In Eqs.~(\ref{fana})-(\ref{rA02}) of the Appendix, we present the analytic expressions for the metric functions and the scalar and vector field profiles corresponding to $N(r)$ given in Eq.~(\ref{Nr}). 
The resulting stellar-type configurations are dynamical in nature and therefore do not correspond to topological defects. As shown in Ref.~\cite{DeFelice:2025vef}, 
the Arnowitt-Deser-Misner (ADM) mass of these ECOs scales as $M\propto r_0$, where $r_0$ is a constant with dimensions of length. Since $r_0$---which is not a parameter of the Lagrangian---can be continuously taken to zero, the configuration disappears smoothly. This behavior demonstrates that the solutions are dynamical rather than topological.

As an extension of the model discussed above, 
we propose the coupling
\be
\mu(\phi)=\mu_0+\mu_1 \frac{\Mpl^p}
{(\phi-\phi_0)^p}\,,
\label{muform}
\ee
where 
\be
p=\frac{2(m+1)}{m}\,,
\ee
and $\mu_0$ and $\mu_1$ are constants, 
while $\phi_0$ denotes the field value 
at $r=0$. Since $m$ is an integer 
in the range $m \ge 2$, the exponent 
satisfies $2 < p \leq 3$.

Near $r=0$, the second term on the right-hand side of 
Eq.~(\ref{muform}) dominates, and the behavior of $\mu(\phi)$ given in Eq.~(\ref{mur=0}) is therefore recovered.
Using Eqs.~(\ref{phia}) and (\ref{mua}) for $N(r)=N_0+N_{2m}r^{2m}$ near $r=0$, the coupling constant $\mu_1$ in Eq.~(\ref{muform}) can be expressed in terms of $N_{2m}$ as 
\be
\mu_1=\frac{q_E^2}{m^2 \Mpl^2} 
\left( \frac{2N_{2m}}{m N_0} \right)^{1/m}\,.
\label{mu1}
\ee
Since $\phi'(r) > 0$ from Eq.~(\ref{rN}), the scalar field increases monotonically with $r$ from its central value $\phi_0$ at $r=0$. As long as $N(r)$ approaches unity at spatial infinity, we have $\phi'(r)\to 0$ as $r \to \infty$, and hence $\phi(r)$ tends to a constant value $\phi_{\infty}$. Consequently, the coupling $\mu(\phi)$ also approaches a constant value, $\mu(\phi_{\infty})$, at spatial infinity.
Because $\phi(r)$ is an increasing function of $r$, $\mu(\phi)$ decreases as $\phi$ increases. For the model with the analytic function (\ref{Nr}) for $N(r)$, however, $\mu(\phi)$ attains a minimum at an intermediate radius and then increases toward the asymptotic value 
$\mu(\phi_{\infty})$ \cite{DeFelice:2025vef}. Therefore, while the two models exhibit similar behavior near $r=0$, their solutions differ at large distances.

\section{Configuration of ECOs}
\label{configsec}

In this section, we derive numerical solutions corresponding to static and spherically symmetric configurations for the models defined by the coupling (\ref{muform}). We recall that the power $p$ is given by $p=2(m+1)/m$, with $m=2,3,4,\cdots$. We introduce the dimensionless ratio
\be
\alpha \equiv 
\frac{\mu_1}{\mu_0}\,.
\label{alphadef}
\ee
Following Ref.~\cite{DeFelice:2024ops}, the absence of ghosts requires the condition $\mu(\phi)>0$. In what follows, we assume $\mu_0>0$ and $\mu_1>0$, so that $\alpha>0$, ensuring that no ghost degrees of freedom appear.

For numerical purposes, it is convenient to obtain approximate solutions for $N$, $h$, and $\phi$ near the origin. For a given value of $m$, the leading-order contribution to $N(r)$ in the expansion (\ref{Nexpan}) is $N(r)=N_0+N_{2m}r^{2m}$.
We introduce a pivot radius $r_0$ and define the following dimensionless quantities:
\ba
\hspace{-0.3cm}
& &
\bar{\phi} \equiv \frac{\phi}{\Mpl},\qquad 
\bar{A}_0 \equiv \frac{A_0}{\Mpl},\qquad
\varphi \equiv \frac{{\rm d}\bar{\phi}}
{{\rm d}x},\qquad
B_0 \equiv \frac{{\rm d}\bar{A}_0}
{{\rm d}x}, \nonumber \\
\hspace{-0.3cm}
& &
x \equiv \frac{r}{r_0},\qquad 
\bar{q}_E \equiv \frac{q_E}{\Mpl r_0},
\ea
and 
\be
\lambda \equiv \frac{m^2}{\bar{q}_E^2} 
\mu_1=r_0^2 \left( \frac{2N_{2m}}{m N_0} 
\right)^{1/m}\,, 
\label{lambda}
\ee
where in the second equality we have 
used the relation (\ref{mu1}).
In terms of these quantities, the leading-order solutions for $N$, $h$, $\bar{\phi}$, $\varphi$, and $B_0$ near $r=0$ can be written as
\ba
N(x) &\simeq& N_0 \left( 1+\frac{m}{2} 
\lambda^m x^{2m} \right)\,,\label{boundary1}\\
h(x) &\simeq& 1-\frac{m^2}{2m+1} 
\lambda^m x^{2m}\,,\\
\bar{\phi}(x) &\simeq& \bar{\phi}_0
+\lambda^{m/2}x^m\,,
\label{boundary3}\\
\varphi(x) &\simeq& m \lambda^{m/2} 
x^{m-1}\,,
\label{boundary4}\\
B_0(x) 
&\simeq& \frac{m^2}{\bar{q}_E}
\sqrt{N_0} \lambda^{m} x^{2m}\,,
\label{boundary5}
\ea
where $\bar{\phi}_0=\phi_0/\Mpl$, and 
we used Eqs.~(\ref{phia}) and (\ref{ha}).

The background equations of motion for 
$N$, $h$, and $\bar{\phi}$ are expressed as
\begin{widetext}
\ba
\hspace{-0.3cm}
\frac{{\rm d}N}{{\rm d}x} &=& 
x \varphi^2 N\,,\label{back1nu}\\
\hspace{-0.3cm}
\frac{{\rm d}h}{{\rm d}x} &=& 
-\frac{\lambda x^2 (h x^2 
\varphi^2+2h-2)({\cal Z}+\alpha)
+\alpha m^2 {\cal Z}}{2\lambda x^3 
({\cal Z}+\alpha)}\,,\label{back2nu} \\
\hspace{-0.3cm}
\frac{{\rm d}^2\bar{\phi}}
{{\rm d}x^2} &=& 
-\frac{[4 \lambda x^3 (h + 1)\varphi 
- m^2 \alpha x \varphi- 2 m(m + 1) \alpha 
{\cal Z}^{-m/(2m + 2)}] \alpha {\cal Z} 
+ [2 \lambda x^2 (h + 1) ({\cal Z}^2 + \alpha^2) 
-\alpha m^2 {\cal Z}^2]x \varphi}
{2 \lambda x^4 h 
({\cal Z}+\alpha)^2}\,,\label{back3nu}
\ea
\end{widetext}
where 
\be
{\cal Z} \equiv \left( \bar{\phi}
-\bar{\phi}_0 \right)^{(2m+2)/m}\,.
\label{calZ}
\ee
The normalized energy density and 
radial pressure, defined as
$\bar{\rho} \equiv \rho r_0^2/\Mpl^2$ and
$\bar{P}_r \equiv P_r r_0^2/\Mpl^2$, 
respectively, can be written as
\be
\bar{\rho}= 
\bar{\rho}_{\phi}+\bar{\rho}_{A_0}\,,
\qquad
\bar{P}_r = 
\bar{\rho}_{\phi}-\bar{\rho}_{A_0}\,,
\ee
where 
\be
\bar{\rho}_{\phi}
=\frac{1}{2}h \varphi^2\,,
\qquad 
\bar{\rho}_{A_0}=\frac{\alpha m^2 {\cal Z}}
{2\lambda ({\cal Z}+\alpha)x^4}\,.
\ee
The normalized transverse pressure is 
given by $\hat{P}_t \equiv P_t\,
r_0^2/\Mpl^2=-\bar{P}_r$. 
We also define the mass function 
\be
{\cal M}(r)=4\pi \Mpl^2 r 
\left[ 1-h(r) \right]\,,
\label{calM}
\ee
which satisfies the differential equation 
${\cal M}'(r)=4\pi \rho r^2$, as follows
from Eq.~(\ref{back1}).
By introducing the dimensionless 
mass function 
\be
\bar{{\cal M}} \equiv 
\frac{{\cal M}}{M_0}\,,\quad 
{\rm where}\quad M_0 \equiv \Mpl^2 r_0\,,
\ee
it follows that 
\be
\frac{{\rm d}\bar{{\cal M}}}{{\rm d} x}
=4\pi \bar{\rho}\,x^2\,.
\ee
The ADM mass $M$ of a compact object 
is known by taking the limit, 
\be
M=\lim_{r \to \infty}{\cal M}(r)
=\lim_{x \to \infty} M_0 
\bar{{\cal M}}(x)\,.
\label{MADM}
\ee

By using Eqs.~(\ref{boundary1})-(\ref{boundary4}), 
which are expanded near $r=0$, the leading-order 
contributions to $\bar{\rho}_{\phi}$ and
$\bar{\rho}_{A_0}$ are given by
\be
\bar{\rho}_{\phi} \simeq 
\bar{\rho}_{A_0} \simeq 
\frac{m^2}{2} \lambda^m 
x^{2m-2}\,.
\ee
Therefore, both the radial pressure $\bar{P}_r$ and the transverse pressure $\bar{P}_t$ vanish at leading order, while the energy density remains positive, $\bar{\rho} \simeq m^2 \lambda^m x^{2m-2}$. This implies that the equations of state, $w_r = \bar{P}_r/\bar{\rho}$ and $w_t = \bar{P}_t/\bar{\rho}$, approach zero near the center, indicating that the scalar and vector fields are in a nonrelativistic regime. This behavior is markedly different from that of standard relativistic stars, 
such as neutron stars composed of baryonic matter.

\begin{figure*}[t]
\centering
\includegraphics[width=0.46\textwidth]{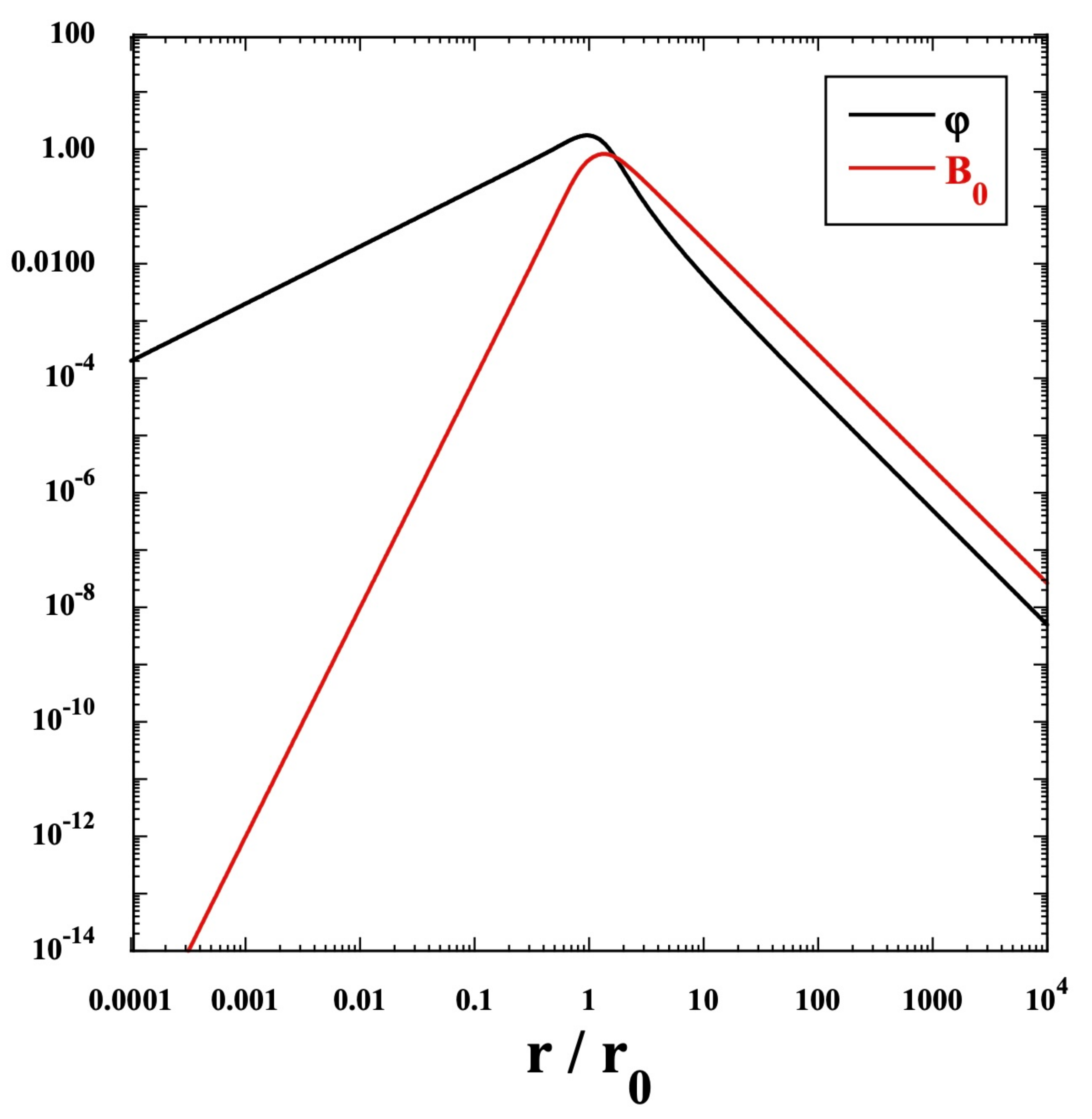}
\hfill
\includegraphics[width=0.46\textwidth]{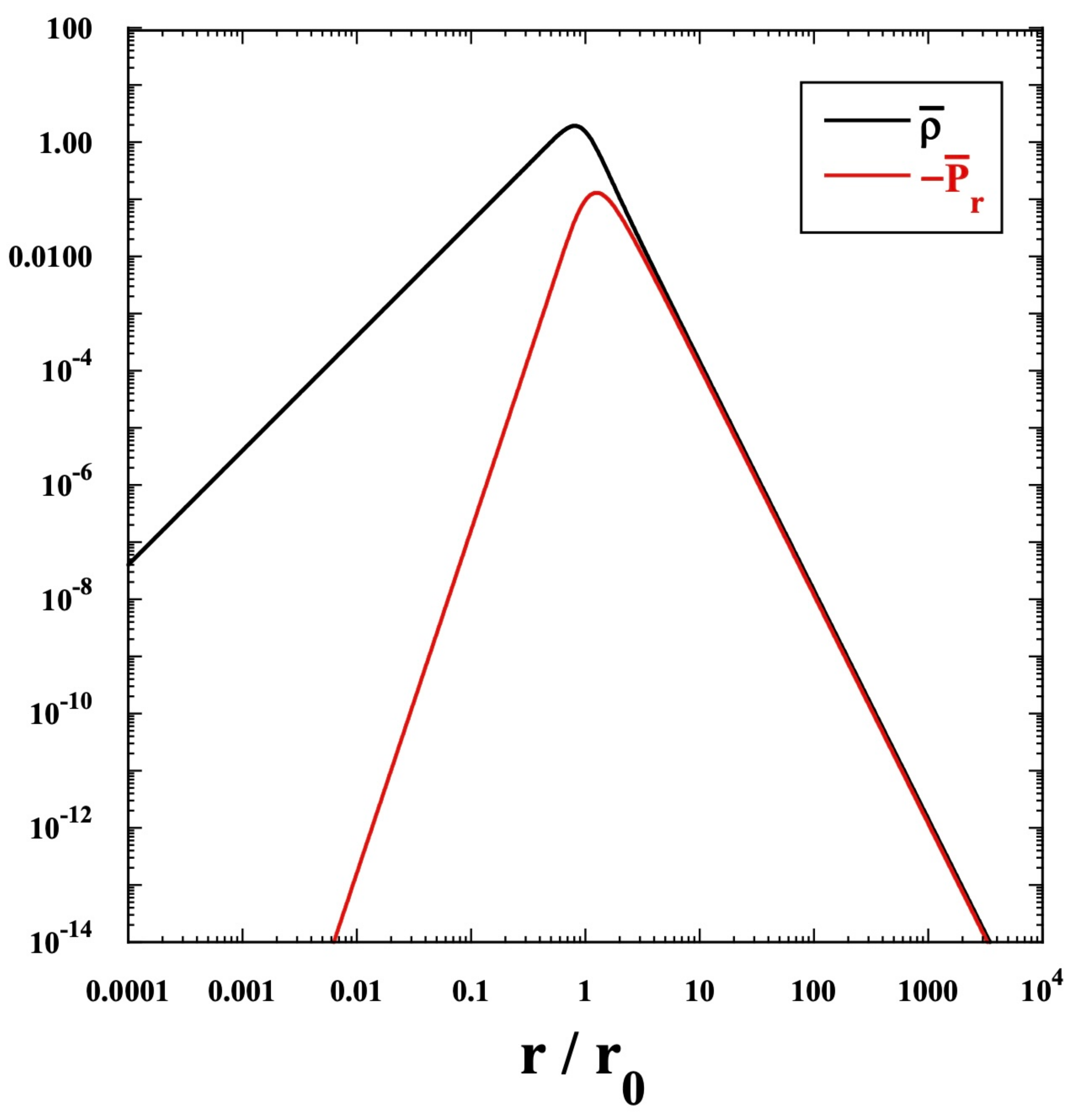}
\caption{The left panel shows the scalar-field 
derivative $\varphi={\rm d}\bar{\phi}/{\rm d}x$ 
and the electric field $B_0 
\equiv {\rm d}\bar{A}_0/{\rm d}x$ as functions of
$x=r/r_0$ for $m=2$, $\lambda=1$, and $\alpha=0.6835$.
The right panel displays $\bar{\rho}$ and $-\bar{P}_r$ 
as functions of $x$ for the same set of parameters. 
We perform the integration outward 
starting from $r = 10^{-4} r_0$.}
\label{fig:profile}
\end{figure*}

Numerically, we integrate the differential 
Eqs.~(\ref{back1nu})-(\ref{back3nu}) outward from a radius close to $r=0$, using the boundary conditions specified in Eqs.~(\ref{boundary1})-(\ref{boundary3}). We note that neither Eq.~(\ref{back2nu}) nor Eq.~(\ref{back3nu}) depends explicitly on the function $N$. From Eq.~(\ref{back1nu}), if $N$ is a solution, then $N/C_0$ is also a solution for any constant $C_0$. We may therefore fix $N_0$ in Eq.~(\ref{boundary1}) 
to 1 (or to any other positive value). 
After obtaining the numerical solution, we rescale it according to $N_0 \rightarrow N_0 / N(r \to \infty)$. This procedure ensures that $N \to 1$ at spatial infinity. We also set $\bar{\phi}_0 = 0$ without loss of generality.

In Fig.~\ref{fig:profile}, we plot $\varphi$ and $B_0$ (left panel), and $\bar{\rho}$ and $-\bar{P}_r$ (right panel), as functions of $x=r/r_0$ for $m=2$, $\lambda=1$, and $\alpha=0.6835$. In the region $r \ll r_0$, the analytic estimates in Eqs.~(\ref{boundary4}) and (\ref{boundary5}) indicate that $\varphi \propto x^{m-1}$ and 
$B_0 \propto x^{2m}$. This behavior agrees with the numerical results for $m=2$, namely $\varphi \propto x$ 
and $B_0 \propto x^4$.
As estimated above, in the regime $x \ll 1$, the energy density grows as $\bar{\rho} \simeq 4\lambda^2 x^2$, while $\bar{P}_r$ remains suppressed relative to $\bar{\rho}$ (see the right panel of Fig.~\ref{fig:profile}). Near $r = r_0$, $\varphi$, $B_0$, $\bar{\rho}$, and $-\bar{P}_r$ all reach their maxima and subsequently decrease for $r \gtrsim r_0$.

As long as the constant term $\mu_0$ provides the dominant contribution to $\mu(\phi)$ in the large-distance region 
($r \gg r_0$), the electric field in Eq.~(\ref{back3}) behaves as $A_0'(r) \simeq q_E/
(\mu_{\infty} r^2) \propto r^{-2}$ with 
$\mu_{\infty}=\mu(\phi_{\infty})$, 
where we have set $N \simeq 1$. 
Taking the limits $\mu_{,\phi} \to 0$, 
$N \to 1$, and $h \to 1$ in Eq.~(\ref{back4}), we similarly obtain the asymptotic behavior $\phi'(r) \simeq q_s/r^2$, 
where $q_s$ is an integration constant. 
In the left panel of Fig.~\ref{fig:profile}, we observe that both $B_0 = {\rm d}\bar{A}_0/{\rm d}x$ and $\varphi = {\rm d}\bar{\phi}/{\rm d}x$ exhibit the asymptotic scaling $B_0 \propto \varphi \propto r^{-2}$. This indicates that the ECOs carry both the electric charge $q_E$ and the scalar charge $q_s$.
For the analytic model in which the metric functions are given in Eqs.~(\ref{fana}) and (\ref{hana}) of Appendix \cite{DeFelice:2024ops,DeFelice:2025vef}, the scalar-field derivative instead behaves as $\varphi \propto r^{-3}$ at large distances, implying the absence of a scalar charge. For the models defined by the couplings in Eq.~(\ref{muform}), we have numerically confirmed that, as long as the ratio $\alpha = \mu_1/\mu_0$ is smaller than unity, the large-distance behavior remains $\varphi \propto r^{-2}$.

As seen in the left panel of Fig.~\ref{fig:profile}, there exists an intermediate region around $r = r_0$ in which $\varphi$ decreases faster than $r^{-2}$. In this regime, the contribution of the term $\mu_1 M_{\rm Pl}^p/(\phi - \phi_0)^p$ in $\mu(\phi)$ is 
non-negligible compared to $\mu_0$.
Since $\varphi$ becomes smaller than $B_0$ for 
$r \gg r_0$, the electric-field energy density 
$\bar{\rho}_{A_0}$ dominates over the scalar-field energy density $\bar{\rho}_{\phi}$. Therefore, in the large-distance regime, we have approximately
$\bar{\rho} \simeq \bar{\rho}_{A_0}$ and
$\bar{P}_r \simeq -\bar{\rho}_{A_0}$,
so that $\bar{\rho} \simeq -\bar{P}_r$. 
As $\phi$ approaches its asymptotic value $\phi_\infty$ in the limit $r \to \infty$, the quantity ${\cal Z}$ in Eq.~(\ref{calZ}) also approaches a constant. Hence, $\bar{\rho}_{A_0} \propto x^{-4}$ in the regime $r \gg r_0$. Indeed, the large-distance behavior $\bar{\rho} \simeq -\bar{P}_r \propto r^{-4}$ is confirmed in the right panel of Fig.~\ref{fig:profile}.

Substituting the large distance solutions  
$\phi'(r) \simeq q_s/r^2$ and 
$A_0'(r) \simeq q_E/(\mu_{\infty} r^2)$ 
into Eq.~(\ref{back1}) and integrating 
it with respect to $r$, the metric 
component $h$ has the following 
asymptotic behavior:
\be
h(r) \simeq 1-\frac{C}{r}
+\frac{q_T^2}{2 \Mpl^2 r^2}\,,
\label{hL}
\ee
where $C$ is an integration constant, 
and
\be
q_T^2 \equiv q_s^2+\frac{q_E^2}
{\mu_{\infty}}\,.
\ee
Using the definition of the ADM mass $M$ 
together with Eq.~(\ref{calM}), 
the constant $C$ is related to $M$ as
\be
C=\frac{M}{4\pi \Mpl^2}\,.
\ee
In addition to the mass $M$, the ECO is endowed
with both scalar and electric charges.

\begin{figure*}[t]
\centering
\includegraphics[width=0.46\textwidth]{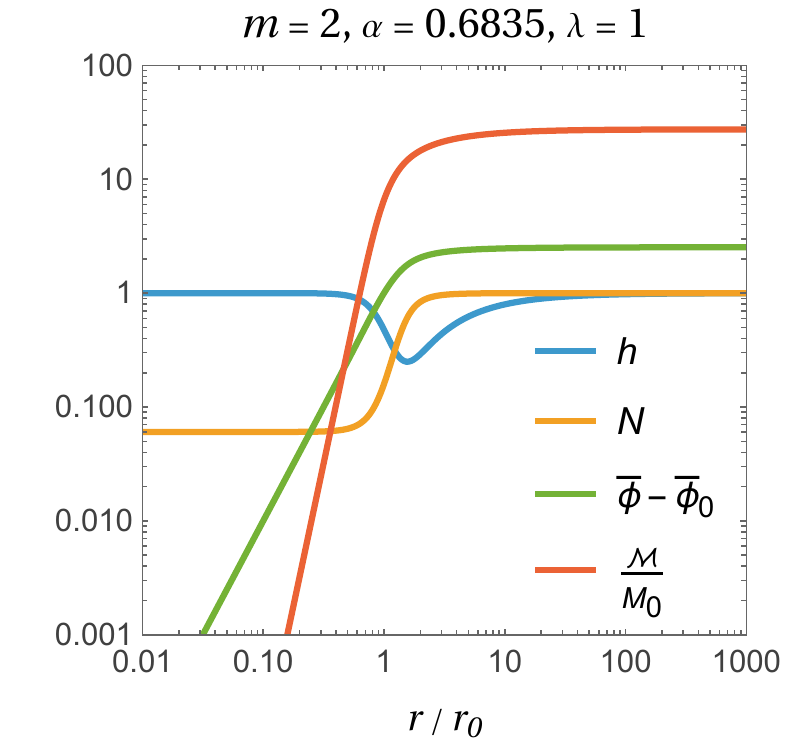}
\hfill
\includegraphics[width=0.46\textwidth]{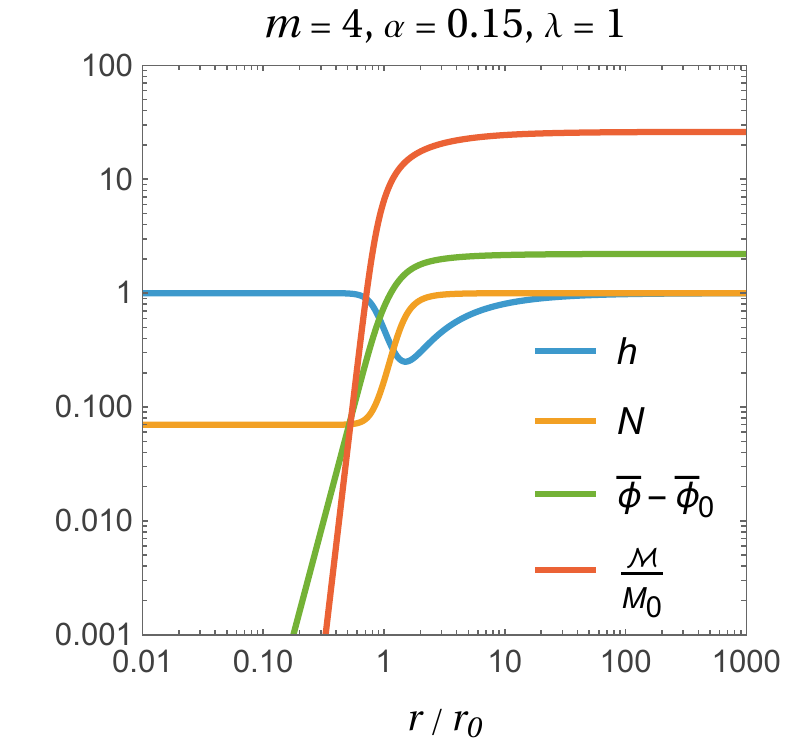}
\caption{Numerical solutions for $h$, $N$, $\bar{\phi}-\bar{\phi}_0$, and ${\cal M}/M_0$ 
as functions of $r/r_0$ for $m=2$ ($p=3$), $\alpha=0.6835$, and $\lambda=1$ (left panel), 
and for $m=4$ ($p=5/2$), $\alpha=0.15$, and $\lambda=1$ (right panel). 
In both panels, we choose values of $\alpha$ slightly smaller than $\alpha_{p}$, where $\alpha_{p}$ denotes the critical value below which light rings are absent. 
The two cases are qualitatively similar, particularly regarding the positions of the minima of $h$. The asymptotic values of $\bar{\phi}$ 
at spatial infinity differ slightly 
between the two cases.
}
\label{fighN}
\end{figure*}

\begin{figure*}[ht]
\centering
\includegraphics[width=0.46\linewidth]{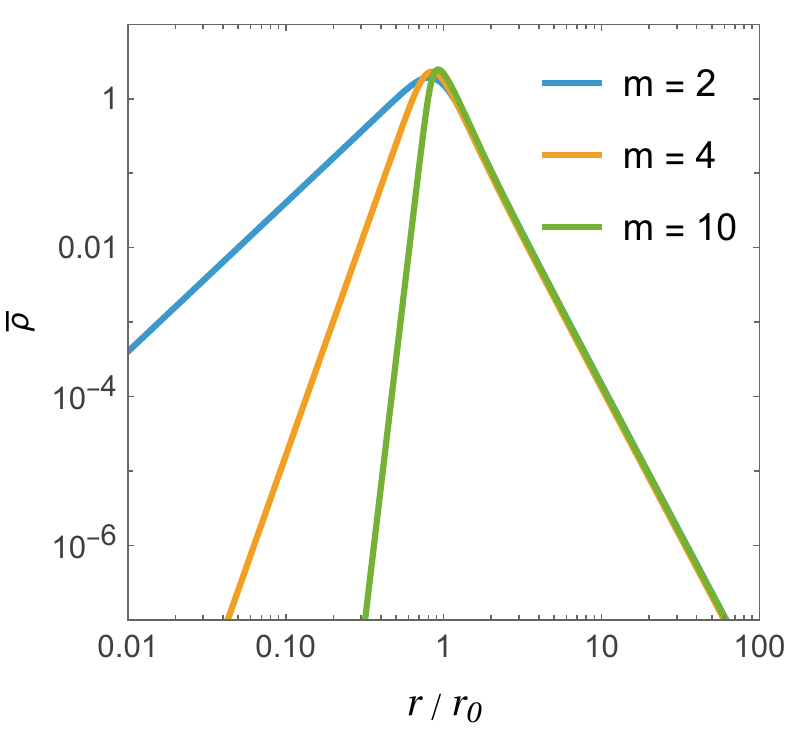}
\includegraphics[width=0.46\linewidth]{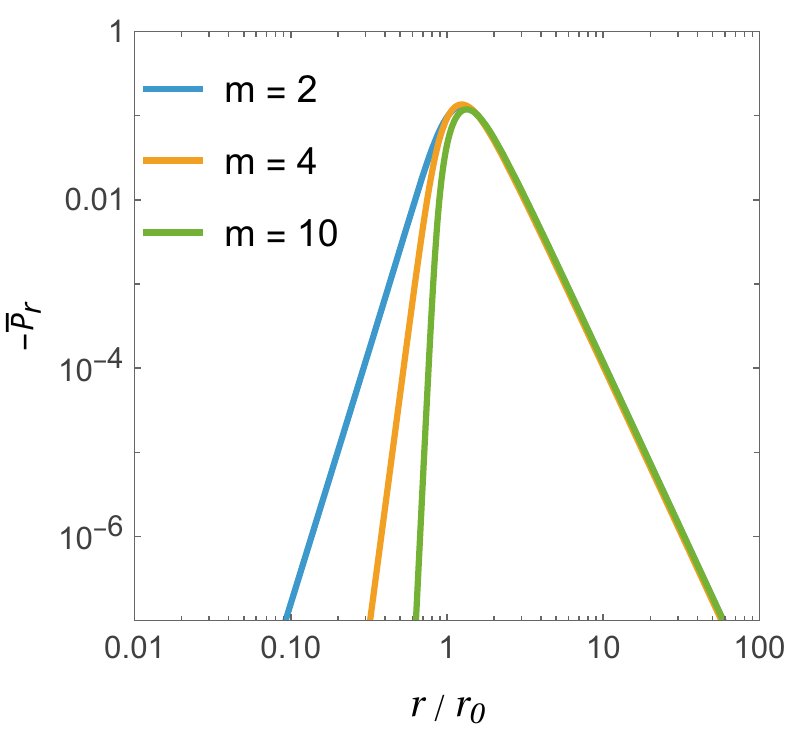}
\caption{Plot of $\bar{\rho}$ (left) and $-\bar{P}_r$ (right) as functions of $r/r_0$ for $m = 2, 4,$ and $10$, with $\lambda = 1$ in all cases. For each value of $m$, we fix $\alpha$ close to $\alpha_p$, namely $\alpha = 0.6835$ for $m = 2$, $\alpha = 0.15$ for $m = 4$, and $\alpha = 0.0261$ for $m = 10$.}
\label{figrhoP}
\end{figure*}

\begin{figure*}[ht]
\centering
\includegraphics[width=0.48\linewidth]{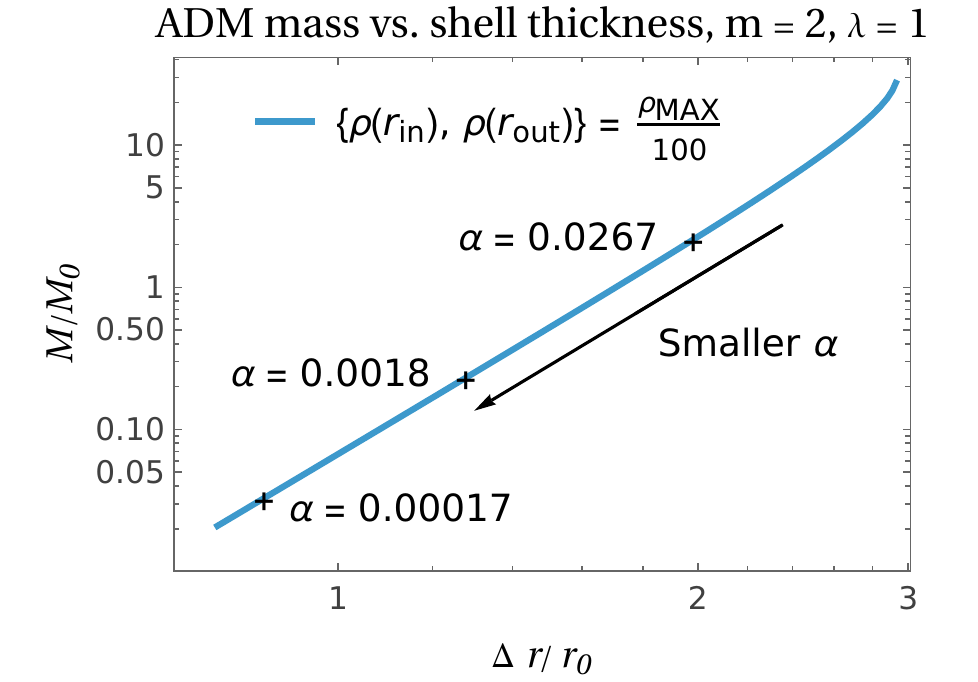}
\includegraphics[width=0.48\linewidth]{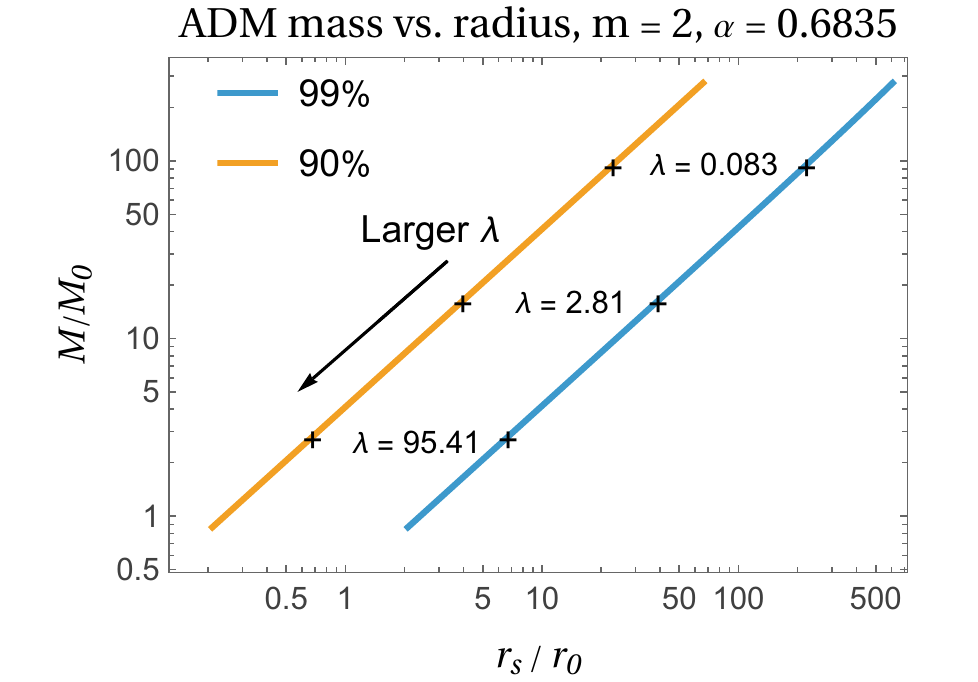}
\caption{
The mass-radius relation for $m=2$ 
is shown. In the left panel, we vary $\alpha$ in the range 
$10^{-5} \leq \alpha \leq 0.6835$ 
while fixing $\lambda=1$, and define the shell thickness 
$\Delta r = r_{\rm out} - r_{\rm in}$, where 
$\rho(r_{\rm out}) = \rho(r_{\rm in}) = \rho_{\rm max}/100$. In the right panel, we instead fix 
$\alpha = 0.6835$ and vary $\lambda$ 
in the range $10^{-2} \leq \lambda \leq 10^3$. 
Here, the radius $r_s$ is defined as the value 
of $r$ satisfying $\mathcal{M}(r_s/r_0)=
(90\%, 99\%)\,M$.
}
\label{figMr}
\end{figure*}

In Fig.~\ref{fighN}, we show $h$, $N$, 
$\bar{\phi}-\bar{\phi}_0$, and ${\cal M}/M_0$ 
as functions of $r/r_0$ for $m=2$,  
$\alpha=0.6835$ (left), and $m=4$, 
$\alpha=0.15$ (right), with $\lambda=1$ 
in both cases. For these model parameters, 
the metric function $h$ attains a positive 
minimum around $r = r_0$ and asymptotically 
approaches 1 in both limits 
$r \ll r_0$ and $r \gg r_0$.
As $\alpha$ increases, the minimum value of $h$ 
approaches 0. On the other hand, 
$h$ must remain positive 
at any distance $r$, as follows from Eq.~(\ref{hr2}).
Therefore, for given values of $m$ and $\lambda$, 
there exists a maximum value of $\alpha$, denoted by $\alpha_{\rm max}$, such that
\be
\alpha<\alpha_{\rm max}\,,
\label{albo1}
\ee
for horizonless, regular ECOs to exist.
As $m$ increases, $\alpha_{\rm max}$ generally 
decreases. For $\lambda = 1$, we find $\alpha_{\rm max} = 1.08$ for $m = 2$ and $\alpha_{\rm max} = 0.23$ for $m = 4$. The values of $\alpha$ adopted in Fig.~\ref{fighN} lie within the range $\alpha < \alpha_{\rm max}$.
In the limit $\alpha \to 0$, i.e., when $\mu(\phi) \to \mu_0 = \text{constant}$, the solution approaches the Minkowski background, with $h(r) = N(r) = 1$ everywhere.
We also find that, for a given $\alpha$, the radius at which $h$ attains its minimum shifts to smaller values as $\lambda$ increases. 

As we will discuss in Sec.~\ref{ringsec}, 
the ECO possesses two light rings
for $\alpha > \alpha_p$, where $\alpha_p$ 
is smaller than $\alpha_{\rm max}$ for same 
values of $m$ and $\lambda$. 
In this case, one of the light rings is 
linearly stable, while the other is linearly unstable.
The single light ring of the Schwarzschild BH 
corresponds to the latter, unstable one.
The linearly stable light ring is nevertheless 
subject to nonlinear instabilities due to 
the accumulation of photon energy along the orbit, 
which can eventually destabilize the ECO.
To avoid such nonlinear instabilities, 
we focus on the coupling $\alpha$ in the range 
\be
\alpha<\alpha_p\,, 
\label{albo2}
\ee
where the upper limit $\alpha_p$ depends 
on $m$ and $\lambda$. 
Since $\alpha_p < \alpha_{\rm max}$ in general, 
the bound (\ref{albo2}) provides a tighter 
constraint than (\ref{albo1}).
The values of $\alpha$ used in the left and right panels of Fig.~\ref{fighN} are both near their respective upper bounds, $\alpha_p$.

In Fig.~\ref{fighN}, the lapse function $N$ increases continuously 
from a value smaller than ${\cal O}(1)$ near $r = 0$ 
to its asymptotic value $1$. 
This behavior is analogous to the analytic function $N(r)$ given by 
Eq.~(\ref{Nr}), where $N(r)$ grows from $N_0~(<1)$ to $1$ 
as a function of $r$ \cite{DeFelice:2024ops,DeFelice:2025vef}. 
The scalar field increases from the value 
$\phi_0$ at $r = 0$ to its asymptotic value 
$\phi_\infty$, with the field derivative 
behaving as 
$\phi'(r) \simeq q_s/r^2$ at large distances.
The mass function ${\cal M}(r)$ also grows 
toward its asymptotic value $M$, 
with the large-distance behavior 
${\cal M}(r) \simeq M - 2\pi q_T^2/r$
following from Eq.~(\ref{hL}).

For the two cases shown in Fig.~\ref{fighN}, 
the ADM masses are 
$M = 27.4\,M_0$ for $m = 2$ 
and $M = 26.1\,M_0$ for $m = 4$, 
exhibiting only a small difference. 
This property persists for larger $m$ and 
values of $\alpha$ close to $\alpha_p$, 
e.g.,  
$M = 27.6\,M_0$ for $m = 10$, 
$\alpha = 0.0261$, and $\lambda = 1$. 
To understand the origin of this behavior, we plot
$\bar{\rho}$ and $-\bar{P}_r$ in Fig.~\ref{figrhoP} 
for $m = 2, 4, 10$ with $\lambda = 1$, 
choosing $\alpha$ close to
$\alpha_p$ for each value of $m$. 
We find that the peak values of $\bar{\rho}$
and the corresponding radii $r_m$
are close to each other across the three cases. 
With increasing $m$, the $\bar{\rho}(r)$ curves 
at $r < r_m$ shift to the right, while those 
at $r > r_m$ show little change. 
A similar behavior is also observed for $-\bar{P}_r$.
The width $\delta r$ containing most of the energy density,
defined, for example, by $\rho(r) > 10^{-6}\,\rho(r_m)$,
is nearly independent of $m$. 
Since the ADM mass of the ECO is roughly proportional 
to $\rho(r_m)\,r_m^2\,\delta r$, the three cases in Fig.~\ref{fighN} exhibit similar values of $M$. 
When $\alpha$ is close to $\alpha_p$, 
the minimum of $h$ is positive but nearly 0.
In such cases, the ADM masses remain similar 
for all $m \geq 2$.

Since the ECO has a shell-like structure with its energy density peaking at $r = r_m$, we define the radius (thickness) of the object as
\be
\Delta r \equiv r_{\rm out} - r_{\rm in}\,,
\label{rs}
\ee
where $r_{\rm out}$ and $r_{\rm in}$ are defined by $\rho(r_{\rm out}) = 0.01\rho(r_m) = \rho(r_{\rm in})$, with $r_{\rm in}<r_{\rm out}$. This definition of the shell thickness reflects the domain-wall--like structure of ECOs. 
In the left panel of Fig.~\ref{figMr}, we show the 
relation between $M/M_0$ 
and $\Delta r/r_0$ for $m=2$ and 
$\lambda=1$ by varying $\alpha$, 
where $M_0$ is related to the pivot 
length scale $r_0$ as $M_0=\Mpl^2 r_0$. 
The compactness of ECOs is defined by 
\be
{\cal C} \equiv \frac{M}{8 \pi \Mpl^2 \Delta r}
=\frac{1}{8\pi} \frac{M}{M_0}\frac{r_0}{\Delta r}\,.
\ee
For $\alpha = 0.6835$, which is close to the upper limit $\alpha_p$, we have $M/M_0 = 27.4$ and $\Delta r/r_0 = 2.93$, yielding a compactness of ${\cal C} = 0.37$.  
The compactness tends to decrease for smaller values of $\alpha$, e.g., ${\cal C} = 0.043$ for $\alpha = 0.0267$.

Instead of $\Delta r$, one can also define the radius $r_s$ as the point where the mass function 
${\cal M}(r)$ reaches 90\,\% or 99\,\% 
of the ADM mass $M$, i.e., ${\cal M}(r_s)=0.90M$ 
or ${\cal M}(r_s)=0.99M$. 
This definition of the radius is adopted in Ref.~\cite{DeFelice:2025vef}. 
In the right panel of Fig.~\ref{figMr}, we plot the relation between $M/M_0$ and $r_s/r_0$ for $m = 2$ and $\alpha = 0.6835$ by varying $\lambda$.  
As $\lambda$ increases, both $M$ and $r_s$ decrease, 
along with a decrease in ${\cal C}$.

In both definitions of the ECO radii, we obtain a compactness in the range 
${\cal O}(10^{-3}) < {\cal C} < {\cal O}(0.1)$, 
whose value depends on the model parameters 
$\alpha$, $m$, and $\lambda$.  
Combined with the peculiar density profile of ECOs, this property allows them to be distinguished from other compact objects, such as neutron stars.

\section{Light rings}
\label{ringsec}

Let us now investigate the geodesic motion of photons in the spacetime generated by the ECOs realized with the coupling (\ref{muform}).  
The existence and stability of light rings provide crucial information about the optical appearance and the dynamical properties of the curved geometry.

\subsection{Existence condition 
for light rings}

To study the trajectories of light rays, 
we write the infinitesimal line element in the form
\begin{equation}
{\rm d}s^2 = 
g_{\alpha\beta}\,{\rm d}x^\alpha {\rm d}x^\beta
= \mathcal{L}\,({\rm d}\lambda)^2 \,,
\end{equation}
where $\mathcal{L}=g_{\alpha\beta}\dot{x}^\alpha \dot{x}^\beta$ and a dot denotes differentiation with respect to the affine parameter $\lambda$. 
The Euler-Lagrange equations 
derived from $\mathcal{L}$,
\begin{equation}
\frac{{\rm d}}{{\rm d}\lambda}
\left( \frac{\partial \mathcal{L}}
{\partial \dot{x}^\mu} \right)
- \frac{\partial \mathcal{L}}{\partial x^\mu} = 0 \,,
\end{equation}
are equivalent to the geodesic equations,
\begin{equation}
\ddot{x}^\alpha +\Gamma^\alpha{}_{\mu \nu}\,
\dot{x}^\mu\,\dot{x}^\nu = 0 \,,
\end{equation}
where $\Gamma^\alpha{}_{\mu\nu}$ denote 
the Christoffel symbols.

For the static, spherically symmetric line element~(\ref{line}) with coordinates $x^\mu=(t,r,\theta,\varphi)$, the Lagrangian takes the form
\begin{equation}
\mathcal{L}
= -f(r)\dot{t}^2
+ h^{-1}(r)\dot{r}^2
+ r^2\dot{\theta}^2
+ (r^2\sin^2\theta) \dot{\varphi}^2 \,.
\label{Lagrangian_photon}
\end{equation}
Variation of ${\cal L}$ with 
respect to $\theta$ yields
\begin{equation}
\ddot{\theta}
+ \frac{2\dot{r}}{r}\dot{\theta}
- (\sin\theta\cos\theta) 
\dot{\varphi}^2 = 0 \,.
\end{equation}
This equation is automatically satisfied for motion confined to the equatorial plane, $\theta = \pi/2$.  
Hence, without loss of generality, we restrict our analysis to equatorial geodesics.

The $\varphi$-component of the geodesic equation,
\begin{equation}
\ddot{\varphi} + 
\frac{2\dot r}{r}\,
\dot{\varphi} = 0 \,,
\end{equation}
admits the first 
integral
\begin{equation}
\dot{\varphi}= 
\frac{L}{r^2} \,,
\label{dotvarphi}
\end{equation}
where $L$ is the conserved angular momentum per unit mass. 
Similarly, the $t$-component of the geodesic 
equation, 
\begin{equation}
\ddot t + \frac{f'}{f}\,\dot r\,\dot t = 0 \,,
\end{equation}
can be integrated to give
\begin{equation}
\dot t = \frac{E}{f} 
\,, \label{dott}
\end{equation}
where $E$ is the conserved energy per unit mass.

Variation of ${\cal L}$ with respect to $r$ 
yields the radial equation
\begin{equation}
\ddot{r}
- \frac{h'}{2h}\,\dot{r}^2
+ h\left(
\frac{E^2 f'}{2f^2}
- \frac{L^2}{r^3}
\right)=0 \,.
\label{radial_photon}
\end{equation}
Since photons follow null geodesics, 
the condition ${\rm d}s^2=0$ leads to 
\begin{equation}
-f(r)\dot{t}^2
+ h^{-1}(r)\dot{r}^2
+ r^2\dot{\varphi}^2 = 0 \,.
\label{null_condition}
\end{equation}
Substituting 
Eqs.~(\ref{dotvarphi}) and 
(\ref{dott})
into~\eqref{null_condition}, we obtain
\begin{equation}
\dot{r}^2
= h(r)\left[
\frac{E^2}{f(r)}
- \frac{L^2}{r^2}
\right] \,.
\label{radial_first_integral}
\end{equation}
Taking the derivative of Eq.~\eqref{radial_first_integral} with respect to $\lambda$ reproduces the second-order radial equation~\eqref{radial_photon}, thereby confirming the consistency of the system.

A light ring corresponds to a circular null orbit at
\begin{equation}
r = r_p = \text{constant} \,,
\end{equation}
for which $\dot{r}=\ddot{r}= 0$. 
From Eq.~\eqref{radial_first_integral}, 
we obtain
\begin{equation}
\frac{E^2}{f(r_p)}
= \frac{L^2}{r_p^2} \,,
\label{constraint1}
\end{equation}
while Eq.~\eqref{radial_photon} gives
\begin{equation}
\left.
\frac{E^2 f'}{2f^2}
\right|_{r=r_p}
=\frac{L^2}{r_p^3} \,.
\label{constraint2}
\end{equation}
Combining Eqs.~\eqref{constraint1} 
and \eqref{constraint2} leads to
\begin{equation}
\left.
(2f - r f')
\right|_{r=r_p}
= 0 \,,
\label{photon_ring_condition}
\end{equation}
which represents the fundamental condition for the existence of light rings.

\begin{figure*}[ht]
\centering
\includegraphics[width=0.45\linewidth]{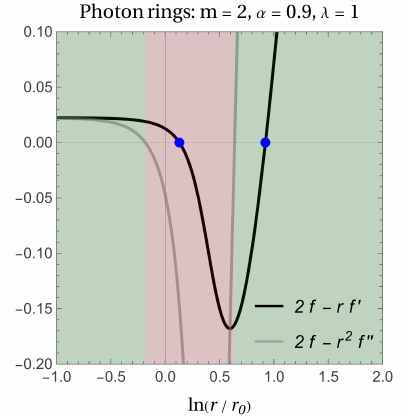}
\includegraphics[width=0.45\linewidth]{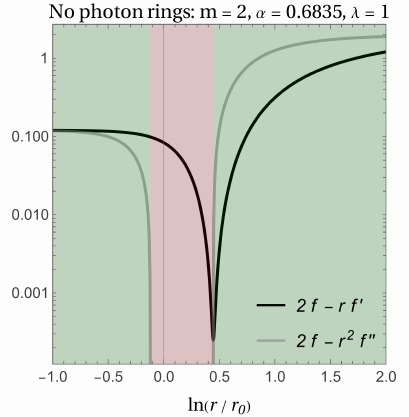}
\caption{
Left panel: Example with two light rings, whose locations correspond to the zeros of $2 f - r f'$ (blue points).  
The shaded regions indicate the sign of $2 f - r^2 f''$: green denotes linearly unstable light rings, while red denotes linearly stable ones.  
Right panel: Example of the marginal case where $\alpha$ is close to the upper limit $\alpha_p$, above which two light rings appear.  
For $\alpha < \alpha_p$, the function $2 f - r f'$ remains positive for all $r$, and no light rings are present.
}
\label{fig:Rings}
\end{figure*}
\subsection{Stability analysis}

To study the radial stability of the null circular orbit, 
we consider a small perturbation 
$\delta_p (\lambda)$ around the orbit,
\begin{equation}
r(\lambda) = r_p\,[1+\delta_p (\lambda)] \,.
\end{equation}
Expanding the radial equation 
(\ref{radial_photon}) to linear order 
in $\delta_p$ and using Eqs.~(\ref{constraint2}) 
and (\ref{photon_ring_condition}) 
to remove $f$ and $f'$ at $r=r_p$, we obtain
\begin{equation}
\ddot{\delta}_p
- \frac{L^2 h(r_p)}
{E^2 r_p^4}
\left[E^2 
- \frac{L^2}{2} f''(r_p)
\right]\delta_p = 0 \,.
\end{equation}
By further exploiting Eq.~(\ref{constraint1}) 
to eliminate $E$, we find
\be
\ddot{\delta}_p-\frac{h(r_p) L^2}
{2f(r_p) r_p^4}
\left( 2f-r^2 f'' \right)|_{r=r_p}
\delta_p =0\,.
\ee
As long as the condition, 
\begin{equation}
\left.
(2f - r^2 f'')
\right|_{r=r_p}
> 0 \,,
\label{instability_condition}
\end{equation}
is satisfied, $\delta_p$ grows as a function of $\lambda$. 
In this case, the light ring is linearly unstable, and perturbations are repelled from the circular orbit.

If instead $(2f - r^2 f'')|_{r=r_p}<0$, 
the light ring is linearly stable.  
The presence of linearly stable light rings generally gives rise to nonlinear instabilities in horizonless compact objects due to the accumulation of energy along photon trajectories \cite{Cunha:2017qtt}.  
Under this condition, there exists one stable light ring in addition to the unstable one.

We denote by $\alpha_p$ the boundary value of $\alpha$ above which two light rings are present. 
For $\alpha<\alpha_p$, no light rings 
are present.
In the left panel of Fig.~\ref{fig:Rings}, which corresponds to the case $\alpha>\alpha_p$, there is a linearly stable ring (the left blue point, satisfying $2f - r^2 f'' < 0$) and a linearly unstable ring (the right blue point, satisfying $2f - r^2 f'' > 0$). In this situation, the accumulation of photon energy along the inner stable trajectory can induce nonlinear instability in the ECOs.
The right panel of Fig.~\ref{fig:Rings} corresponds to the marginal case, in which $\alpha$ is close to $\alpha_p$. In this case, the condition $2f - r f' = 0$ admits no solution, and consequently, no light rings are present. The absence of linearly stable light rings indicates that the ECOs are not prone to nonlinear 
instability. Therefore, the parameter $\alpha$ should be restricted to the range 
\be
\alpha<\alpha_p\,.
\ee
Indeed, in the numerical simulations presented in Sec.~\ref{configsec}, this bound on $\alpha$ has already been incorporated.
Note that $\alpha_p$ is generally smaller than $\alpha_{\rm max}$, where $\alpha_{\rm max}$ denotes the marginal value of $\alpha$ below which the ECOs can exist.

\subsection{Absence of photon echoes}

We are also interested in whether photons propagating into the ECOs give rise to echoes.
For this purpose, we treat the photon as a test particle characterized by the electromagnetic field strength $\tilde{F}_{\mu \nu}
=\partial_{\mu} \tilde{A}_{\nu}-\partial_{\nu} \tilde{A}_{\mu}$, 
where $\tilde{A}_{\mu}$ denotes the 
electromagnetic four-potential.
The dynamics of the photon are then 
governed by the Maxwell equations,
\begin{equation}
\nabla_\nu 
\tilde{F}^{\mu\nu} = 0 \,.
\end{equation}
It is well known that, upon decomposing the vector potential $\tilde{A}_\mu$ into axial (odd) and polar (even) modes, the perturbation equations reduce to a single Schr\"odinger-like equation for the two independent photon polarizations $\psi_i$ ($i=1,2$); see, e.g., Refs.~\cite{Berti:2009kk,Nakarachinda:2025bvy}. 
Remarkably, both polarization states satisfy the same master equation, which takes the form
\begin{equation}
\frac{{\rm d}^2 \psi_i}
{{\rm d} r_*^2}+
\left[ \omega^2 - V(r) \right]
\psi_i = 0 \,,
\end{equation}
where $r_*=\int (fh)^{-1/2}{\rm d}r$ 
is the tortoise coordinate, and 
$V(r)$ is the effective potential 
defined by 
\begin{equation}
V(r) = \frac{l(l+1)}{r^2} f(r) \,,
\end{equation}
with $l$ denoting the multipole 
number.

The occurrence of photon echoes is associated with the presence of at least one local minimum of $V(r)$, which would give rise to a trapping region between potential barriers. 
A necessary condition for the existence 
of such a minimum is
\begin{equation}
\frac{{\rm d}V}{{\rm d}r}
=\frac{l(l+1)}{r^3}
\left(
r f' - 2 f
\right)= 0 \,,
\end{equation}
or, equivalently, ${\rm d}V/{\rm d}r_* = 0$ 
for some value of $r$. 
The condition for the existence of light rings coincides with the vanishing of the function $r f' - 2 f$. 
Therefore, if light rings are absent, the quantity $r f' - 2 f$ never vanishes, implying that ${\rm d}V/{\rm d}r \neq 0$ for all $r$. 
In this case, the effective potential does not develop any local minimum, and no trapping region is formed. 
Consequently, the absence of light rings 
is equivalent to the absence of 
electromagnetic echoes.

\section{Gravitational lensing}
\label{lensingsec}

%
\begin{figure*}[ht]
\centering
\includegraphics[width=0.45\linewidth]{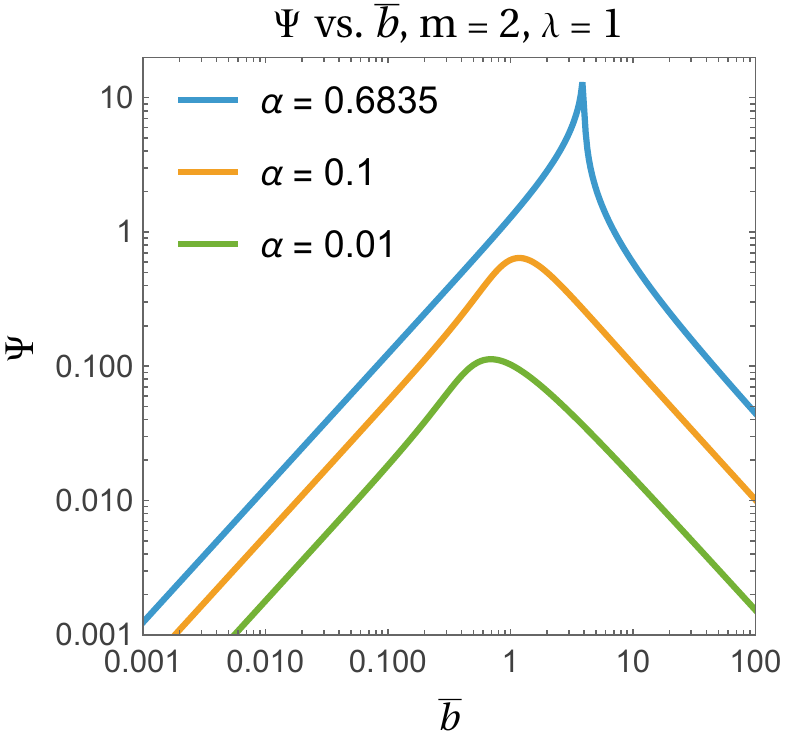}
\includegraphics[width=0.45\linewidth]{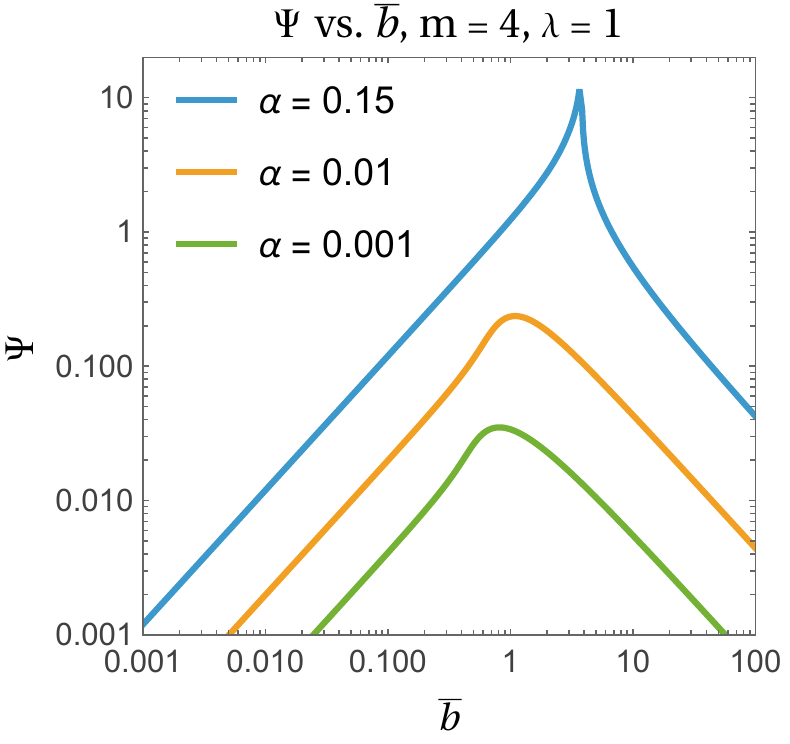}
\caption{Deflection angle $\Psi$ as a function 
of $\bar b = b/r_0$.
Left panel: Numerical evaluation for $\alpha = 0.6835, 0.1, 0.01$ with $m = 2$ and $\lambda =1$.
Right panel: Numerical evaluation for 
$\alpha = 0.15, 0.01, 0.001$ with 
$m = 4$ and $\lambda = 1$.
The values $\alpha = 0.6835$ and $\alpha = 0.15$ are
close to $\alpha_p$, above which light rings form.  
For such marginal values of $\alpha$, photon orbits remain unstable, yet the deflection angle can exceed $2\pi$, allowing photons to complete multiple revolutions before escaping. The precise maximum of $\Psi(\bar b)$ 
is numerically delicate, 
as it depends sensitively on the integration range of differential equations, solver method, and required precision. Near the peak, however, we consistently find $\Psi = \mathcal{O}(10)$ for both $m=2$ and $m=4$. 
Due to these numerical uncertainties, we do not show the exact maximum, but only values of $\bar b$ near it for which $\Psi(\bar b) \simeq 10$.
\label{lensingfig}
}
\end{figure*}

To study the gravitational lensing induced by the ECOs, we consider a photon trajectory confined to the 
$\theta = \pi/2$ plane, as in Sec.~\ref{ringsec}. 
We can rewrite Eq.~(\ref{null_condition}) 
in the form
\be
\dot{r}^2=hE^2 \left( \frac{1}{f}
-\frac{b^2}{r^2} \right)\,,
\ee
where $b=L/E$ is the impact parameter. 
By using the relation $\dot{\varphi}=L/r^2$, 
we obtain the following relation 
\be
\frac{{\rm d} \varphi}{{\rm d}r}
= \pm \frac{1}{r \sqrt{h}} 
\left( \frac{r^2}{b^2 f} - 1 \right)^{-1/2}\,.
\ee
The distance $r_c$, at which the photon approaches 
the ECO most closely, corresponds to the point 
where ${\rm d}r/{\rm d}\varphi = 0$.  
This condition yields
\be
b^2 = \frac{r_c^2}{f(r_c)}\, .
\label{bs}
\ee

The total deflection angle of 
the light ray traveling from 
one side, passing at $r=r_c$, 
and moving toward the 
opposite side is 
\be
\varphi_{\rm total}=2 \int_{r_c}^{\infty}
\frac{1}{r \sqrt{h}} \left[ \frac{r^2}{r_c^2} 
\frac{f(r_c)}{f(r)}-1 \right]^{-1/2} {\rm d}r\,,
\label{varphi}
\ee
where we have substituted Eq.~(\ref{bs}). 
On the Minkowski background, where both 
$f(r)$ and $h(r)$ are equal to 1, 
the angle (\ref{varphi}) reduces to $\pi$. 
On a curved background, the deflection angle 
$\Psi$ is then given by 
\be
\Psi = 2 \int_{r_c}^{\infty}
\frac{1}{r \sqrt{h}} 
\left[ \frac{r^2}{r_c^2} 
\frac{f(r_c)}{f(r)} 
- 1 \right]^{-1/2} {\rm d}r 
- \pi \,.
\label{Psi}
\ee

For a given impact parameter $b$, the corresponding 
value of $r_c$ can be obtained from Eq.~(\ref{bs}).  
The deflection angle is then determined 
by evaluating the integral in Eq.~(\ref{Psi}). 
Since photons interact with the ECOs composed of scalar and vector fields in the dark sector only through gravity, the impact parameter can be arbitrarily small. 
In the limit $b \to \infty$, where the spacetime approaches Minkowski, we expect $\Psi \to 0$.  
A similar behavior is expected in the limit $b \to 0$, since the spacetime near $r = 0$ corresponds to 
a weak-gravity regime with suppressed energy 
density and pressure.

In Fig.~\ref{lensingfig}, we show $\Psi$ as a function of $\bar{b} = b/r_0$ for (i) $m=2$ and $\lambda=1$ with $\alpha = 0.6835, 0.1, 0.01$ (left panel), and (ii) $m=4$ and $\lambda=1$ with $\alpha = 0.15, 0.01, 0.001$ (right panel). For each value of $\alpha$, $\Psi$ exhibits a peak at $\bar{b}$ of order 0.1--1.
Since the background spacetime is close to 
Minkowski in the limits $r \ll r_0$ and $r \gg r_0$, 
the deflection angle approaches 0 
for $|\bar{b}| \ll 1$ and $\bar{b} \gg 1$.  
The deviation from Minkowski spacetime is largest 
for $r \sim r_0$, so that $\Psi$ attains 
its maximum at intermediate values of $\bar{b}$.

As shown in the left panel of Fig.~\ref{lensingfig} (the $m=2$ case), the maximum values of $\Psi (\bar{b})$ for $\alpha = 0.01$ and $\alpha = 0.1$ are $\Psi = 0.11$ 
at $\bar{b}=0.66$ and 
$\Psi = 0.64$ at 
$\bar{b}=1.2$, respectively. 
For $\alpha = 0.6835$, the deflection angle can reach $\mathcal{O}(10)$, in which case the photon completes more than one orbit. 
When $\alpha$ is close to $\alpha_p$, the numerical integration of Eq.~(\ref{Psi}) becomes unstable near the maximum of $\Psi(\bar{b})$. 
Therefore, for the case $\alpha = 0.6835$ in Fig.~\ref{lensingfig}, we omit plotting the values of $\Psi$ in the vicinity of its maximum. 
For $\alpha = 0.6835$, we also observe that the peak of $\Psi(\bar{b})$ around $\bar{b} = 4.0$ is sharper than those for $\alpha = 0.1$ and $\alpha = 0.01$.  
As $\alpha$ increases toward $\alpha_p$, the magnitude of $\Psi(\bar{b})$ generally increases, with the peak shifting toward larger values of $\bar{b}$.

In the right panel of Fig.~\ref{lensingfig} (the $m=4$ case), we observe a behavior of $\Psi(\bar{b})$ 
similar to that in the left panel.
In this case, the maximum 
values of $\Psi (\bar{b})$ 
realized for $\alpha=0.001$ 
and $\alpha=0.01$ are 
$\Psi (\bar{b})=0.036$ 
at $\bar{b}=0.79$ and
$\Psi (\bar{b})=0.25$ 
at $\bar{b}=1.1$. 
As $\alpha$ increases, $\Psi(\bar{b})$ grows in amplitude, and its peak shifts toward larger $\bar{b}$. 
For $\alpha = 0.15$, $\Psi(\bar{b})$ reaches $\mathcal{O}(10)$ near its maximum at $\bar{b} = 3.8$. 
The presence of peaks in $\Psi(\bar{b})$ at $\bar{b} \sim 0.1$--$1$ constitutes a distinctive signature 
of ECOs with a shell-like structure.

In the Appendix, we also compute the deflection angle for the analytic model specified by the metric functions (\ref{fana}) and (\ref{hana}).
Analogous to the discussion above, $\Psi$ exhibits maxima for impact parameters $b$ of order $r_0$.
The peak values of $\Psi$ increase as $N_0$ decreases toward its lower bound, which is set by the requirement that no light rings exist.

\section{Innermost stable circular orbits}
\label{ISCOsec}

Finally, we investigate the possibility of the existence of ISCOs for our ECOs. 
To this end, we analyze the geodesic motion of massive particles in their vicinity. 
Analogous to the photon case, we consider the trajectories of particles confined to the equatorial plane, $\theta = \pi/2$, without loss of generality.  
The conservation of angular momentum $L$ and energy $E$ leads to equations analogous to Eqs.~(\ref{dotvarphi}) and (\ref{dott}), namely
\be
\dot{\varphi}=\frac{L}{r^2}\,,
\qquad 
\dot{t}=\frac{E}{f}\,,
\ee
where a dot denotes differentiation 
with respect to the conformal time $\tau$.

For massive particles, the normalization 
condition for the four-velocity,
\begin{equation}
g_{\mu\nu}\,\dot x^\mu \dot x^\nu = -1\,,
\end{equation}
yields
\begin{equation}
\frac{\dot r^{\,2}}{h} + \frac{L^2}{r^2} 
- \frac{E^2}{f} + 1 = 0 \,.
\label{eq:normalization}
\end{equation}
The radial geodesic equation 
takes the same form as 
Eq.~(\ref{radial_photon}), i.e., 
\begin{equation}
\ddot{r}
- \frac{h'}{2h}\,\dot{r}^2
+ h \left(
\frac{E^2 f'}{2f^2}
- \frac{L^2}{r^3}
\right)=0  \,.
\label{eq:radial_geodesic}
\end{equation}

\begin{figure*}[ht]
\centering
\includegraphics[width=0.45\linewidth]{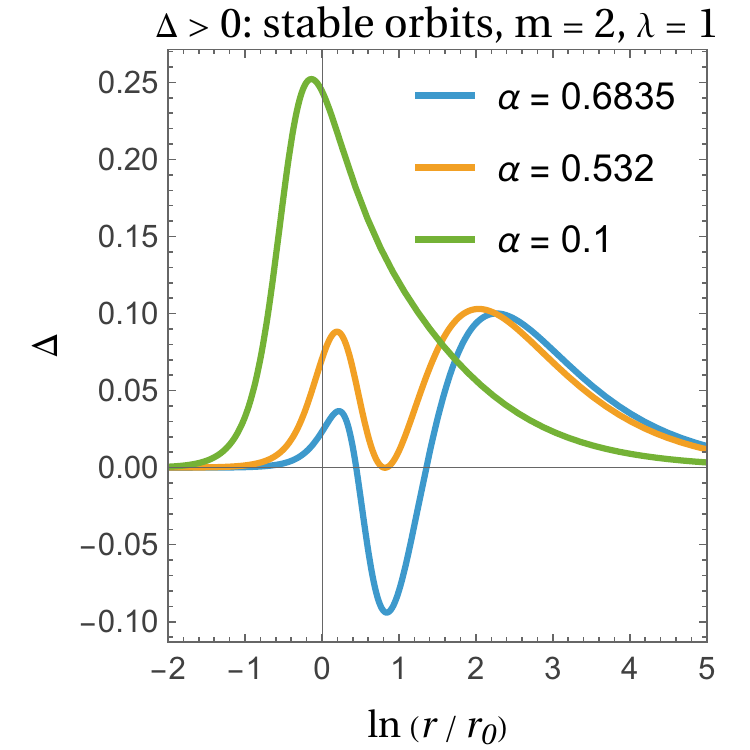}
\includegraphics[width=0.45\linewidth]{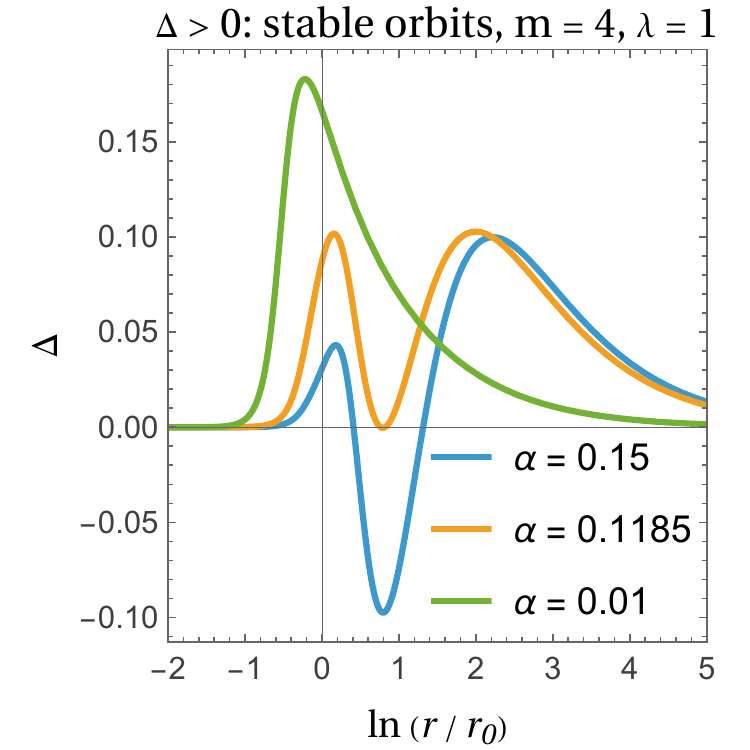}
\caption{
The quantity $\Delta$ as a function of $r/r_0$ for (i) $\lambda = 1$, $m = 2$ (left panel) and (ii) $\lambda = 1$, $m = 4$ (right panel). Stability of circular orbits of massive particles requires $\Delta > 0$. The critical values of $\alpha$ above which solutions to $\Delta(r)=0$ exist are $\alpha_{\mathrm{ISCO}} = 0.5320$ (left) and $\alpha_{\mathrm{ISCO}} = 0.1185$ (right). In the finite region where $\Delta(r) < 0$, circular orbits are unstable.
}
\label{fig:Delta}
\end{figure*}

Let us now consider a circular orbit 
characterized by $r = r_1 = \mathrm{const}$.  
In this case, $\dot{r} = \ddot{r} = 0$, and hence Eqs.~(\ref{eq:normalization}) and (\ref{eq:radial_geodesic}) reduce, respectively, to
\ba
& &
\frac{L^2}{r_1^2} + 1 = 
\frac{E^2}{f(r_1)} \,,
\label{cir1} \\
& &
h(r_1) \left[
\frac{E^2 f'(r_1)}{2f(r_1)^2}
-\frac{L^2}{r_1^3}
\right]=0\,, 
\label{cir2}
\ea
From Eq.~(\ref{cir1}), 
it follows that $f(r_1) > 0$.
Combining Eqs.~(\ref{cir1}) and (\ref{cir2}), 
we obtain
\begin{equation}
h(r_1)\left[
\frac{E^2 f'(r_1)}{2 f(r_1)^2}
-\frac{E^2-f(r_1)}{r_1 f(r_1)}
\right] = 0 \,.
\end{equation}
Assuming that $h(r_1)>0$, this equation 
leads to
\begin{equation}
E^2 = \frac{2 f(r_1)^2}{2 f(r_1) - r_1 f'(r_1)} \,,
\label{Es}
\end{equation}
which in turn requires
\begin{equation}
2 f(r_1) - r_1 f'(r_1) > 0 \,.
\label{conf}
\end{equation}
Substituting Eq.~(\ref{Es}) into 
Eq.~(\ref{cir1}), the corresponding angular 
momentum squared is given by 
\begin{equation}
L^2 = \frac{r_1^3 f'(r_1)}
{2 f(r_1) - r_1 f'(r_1)} \,.
\label{Ls}
\end{equation}
Along with the condition (\ref{conf}), 
the non-negativity of $L^2$ requires that 
$f'(r_1)$ should be in the range
\begin{equation}
0 \le f'(r_1) < \frac{2 f(r_1)}{r_1} \,.
\end{equation}

To study the stability of the circular orbit,
we introduce a small radial perturbation $\delta$
around $r = r_1$ as
\begin{equation}
r(\tau) = r_1 + \delta(\tau)\,.
\end{equation}
Substituting Eqs.~(\ref{Es}) and~(\ref{Ls}) into
Eq.~(\ref{eq:radial_geodesic}) and expanding
Eq.~(\ref{eq:radial_geodesic}) to linear order 
in $\delta$, we obtain
\begin{equation}
\ddot{\delta} 
+ \frac{h(r_1)\,\Delta(r_1)}
{r_1^2 f(r_1)\left[ 2 f(r_1) 
- r_1 f'(r_1) \right]} 
\, \delta = 0 \,,
\label{deleq}
\end{equation}
where
\begin{equation}
\Delta(r_1)
\equiv r_1^2 f(r_1) f''(r_1)
- 2 r_1^2 [f'(r_1)]^2
+ 3 r_1 f(r_1) f'(r_1) \,.
\end{equation}
The stability of the circular orbit requires that the coefficient of $\delta$ in Eq.~(\ref{deleq}) 
be positive, which reduces to the condition
\begin{equation}
\Delta(r_1) > 0 \,.
\end{equation}
The radius of the ISCO, $r_{\rm ISCO}$, 
is determined by the condition 
\be
\Delta(r_{\rm ISCO})=0\,.
\ee
As long as solutions to this equation exist, 
the ECOs can possess an ISCO.

In Fig.~\ref{fig:Delta}, 
we plot the quantity 
\be
\Delta (r)= r^2 f(r) f''(r)-2 r^2 [f'(r)]^2
+ 3 r f(r) f'(r)\,,
\ee
as a function of $r$ for two different 
values of $m$ with $\lambda=1$.
There are two solutions to $\Delta(r)=0$ 
for $\alpha > \alpha_{\mathrm{ISCO}}$, where $\alpha_{\mathrm{ISCO}} = 0.5320$ for $m=2$, $\lambda=1$ (left panel) and $\alpha_{\mathrm{ISCO}} = 0.1185$ for $m=4$, $\lambda=1$ (right panel). 
Note that the values $\alpha = 0.6835$ (left) and $\alpha = 0.15$ (right), chosen in Fig.~\ref{fig:Delta}, are both close to $\alpha_p$, above which two light rings appear.
We denote the two solutions of $\Delta(r)=0$ by $r_{\mathrm{ISCO,min}}$ and $r_{\mathrm{ISCO,max}}$, with $r_{\mathrm{ISCO,min}} < r_{\mathrm{ISCO,max}}$.
For distances $r$ in the range
\be
r_{\rm ISCO,min}<r<r_{\rm ISCO,max}\,,
\label{rrange}
\ee
we have $\Delta (r)< 0$, meaning that circular orbits within this interval are radially unstable.
Conversely, stable circular motion with 
$\Delta(r)>0$ is allowed for
$r < r_{\rm ISCO,min}$ and for 
$r > r_{\rm ISCO,max}$.

The Schwarzschild BH, which is characterized 
by the metric function 
$f = 1 - 2M/r$, gives the value 
$\Delta(r) = 2M (r - 6M)/r^2$, 
so that the ISCO is located at 
$r_{\rm ISCO} = 6M$.
In this case, we have $\Delta(r) < 0$ for 
$r < r_{\rm ISCO}$, whereas 
$\Delta(r) > 0$ for $r > r_{\rm ISCO}$.  
This behavior differs from that of 
$\Delta(r)$ around $r_{\rm ISCO, min}$ 
for our ECOs, where $\Delta(r) > 0$ for 
$r < r_{\rm ISCO, min}$ and 
$\Delta(r) < 0$ for 
$r_{\rm ISCO, min} < r < r_{\rm ISCO, max}$.
Near $r = r_{\rm ISCO, max}$, $\Delta(r)$ is 
negative for $r$ smaller than 
$r_{\rm ISCO, max}$. 
These results show that circular orbits tend to be unstable in the intermediate regime (\ref{rrange}), where most of the ECO density is concentrated, whereas they are stable at small radii, 
$r < r_{\mathrm{ISCO,min}}$.
This property is different from the 
Schwarzschild BH, where circular orbits 
are unstable in the region near the center 
($r < r_{\rm ISCO}$).

The critical value $\alpha_{\mathrm{ISCO}}$ depends on the model parameters $m$ and $\lambda$. For fixed $\lambda$, $\alpha_{\mathrm{ISCO}}$ tends to decrease as $m$ increases. 
As in the cases shown in Fig.~\ref{fig:Delta}, $\alpha_{\mathrm{ISCO}}$ is generally smaller than $\alpha_p$.
Therefore, in the parameter range
\be
\alpha_{\mathrm{ISCO}}<\alpha<\alpha_p\,,
\ee
ISCOs are present, while no light rings exist.
For $\alpha < \alpha_{\mathrm{ISCO}}$, we have $\Delta > 0$ for all $r > 0$. In this regime, no ISCOs are present, and all circular orbits are stable.

\section{Conclusions} 
\label{consec}

In this paper, we have constructed new asymptotically flat ECO solutions endowed with both scalar and electric charges and investigated their observational consequences within the class of ESM theories described by the action (\ref{action}).
These ECOs are supported by scalar and vector fields in the dark sector, which interact through the coupling $\mu(\phi) F$. Unlike previous related works \cite{DeFelice:2024ops,DeFelice:2025vef}, where $\mu(\phi)$ is reconstructed to reproduce analytic metric functions, we have proposed an explicit coupling of the form (\ref{muform}) for realizing the ECOs.

As shown in Ref.~\cite{Herdeiro:2019iwl}, the coupling $\mu(\phi)$ must diverge to ensure regular boundary conditions at the center.
This divergence corresponds to a weak-coupling 
regime of the theory and therefore does not signal any physical pathology.
Indeed, all physical quantities---including the energy density, pressure, and the scalar and vector field profiles---remain finite everywhere.
Moreover, neither ghost nor Laplacian instabilities arise, owing to the positivity of $\mu(\phi)$ at all radii $r$.
Near $r=0$, the analytic model proposed in Refs.~\cite{DeFelice:2024ops,DeFelice:2025vef} exhibits the behavior $\mu(\phi) \propto (\phi-\phi_0)^{-3}$, which mimics the coupling (\ref{muform}) with $m=2$.
While the analytic model features a minimum of $\mu(\phi)$ at an intermediate radius, the coupling proposed in this paper decreases monotonically as a function of $r$ and approaches a constant value $\mu_{\infty}$ at spatial infinity. 

The density profile of the resulting ECO exhibits the remarkable feature that $\rho(r)$ attains a maximum at an intermediate radius. In particular, it vanishes in both limits $r \to 0$ and $r \to \infty$.
This behavior naturally gives rise to configurations resembling spherical shells, with the density effectively concentrated within a finite radial region.
We have computed the radius (thickness) of the ECO, $\Delta r$ defined in Eq.~(\ref{rs}), as well as the ADM mass, taking into account its 
domain-wall--like structure.
While the compactness of the object, ${\cal C}$, depends on the model parameters, it can reach values as large as ${\cal O}(0.1)$ when the parameter $\alpha = \mu_1/\mu_0$ approaches its upper bound $\alpha_p$, which is constrained by the absence of light rings.

On the phenomenological side, we have shown that ECOs with a peculiar density profile can affect observable quantities associated with null and timelike geodesics. We first investigated the structure of light rings and demonstrated that two distinct light rings arise for $\alpha > \alpha_p$, where $\alpha_p$ depends on the other model parameters $m$ and $\lambda$. In particular, one of them is linearly stable, whereas the other is unstable. Since the linearly stable light ring may be subject to nonlinear instabilities due to the accumulation of photon energy, we have constrained the parameter space to $\alpha < \alpha_p$. Under this condition, we further showed that photons plunging into the ECO do not exhibit echoes, owing to the absence of a local minimum in the effective potential.

We have also computed the gravitational-lensing deflection angle $\Psi$ as a function of the impact parameter $b$. We found that $\Psi$ exhibits peaks at intermediate values of $b$, while $\Psi \to 0$ in the limits $b \to 0$ and $b \to \infty$. This behavior is intrinsically related to the shell-like structure of the ECO, 
whose density vanishes as $r \to 0$ and $r \to \infty$. 
For fixed values of $m$ and $\lambda$, the deflection 
angle increases with $\alpha$.
In particular, when $\alpha$ is close to $\alpha_p$, the maximum value of $\Psi$ can reach $\mathcal{O}(10)$.

We also investigated the ISCOs of massive particles and 
showed that they exist for $\alpha_{\rm ISCO} < \alpha < \alpha_p$, where the lower bound $\alpha_{\rm ISCO}$ depends on the model parameters $m$ and $\lambda$. Within this range of $\alpha$, circular orbits are unstable in the radial interval $r_{\rm ISCO, min} < r < r_{\rm ISCO, max}$, whereas they are stable for $r < r_{\rm ISCO, min}$ and $r > r_{\rm ISCO, max}$. For $\alpha < \alpha_{\rm ISCO}$, no ISCOs exist, and all circular orbits are stable.

The correlated behavior of light rings, gravitational-lensing observables, and ISCOs---which exhibit more significant deviations from the Minkowski background for larger $\alpha$---reinforces the predictive coherence of the framework. ECOs with a shell-like structure leave characteristic imprints on strong-field observables, thereby providing concrete avenues for observational discrimination 
from other compact objects.

Future work should aim to place constraints on the model parameters using high-precision gravitational-lensing and horizon-scale imaging observations. Observations of gravitational waves from compact binaries containing ECOs can further constrain the scalar and electric charges through modifications of the waveform during the inspiral phase \cite{Will:1994fb,Alsing:2011er,Liu:2020moh,Wang:2020ori,Niu:2021nic,Higashino:2022izi,Takeda:2023wqn}. Moreover, gravitational-wave observations can provide information on the tidal deformability of ECOs, thereby placing constraints on their internal structure \cite{Hinderer:2007mb,LIGOScientific:2017vwq,De:2018uhw,Diedrichs:2025vhv}. Analyses of current and upcoming data should allow us to place upper bounds on the dark charges of ECOs. It is also of interest to construct solutions describing rotating ECOs and to investigate their observational signatures. These issues are left for future work.

\section*{Acknowledgements}

We thank Vitor Cardoso for useful discussions and correspondence. We are also grateful to Torben Frost for valuable correspondence.
ST is supported by JSPS KAKENHI Grant Number 22K03642 and Waseda University Special Research 
Projects (Nos.~2025C-488 and 2025R-028).

\section*{Appendix:~Gravitational lensing 
for exact solutions}
\renewcommand{\theequation}{A.\arabic{equation}} 
\setcounter{equation}{0}
%

\begin{figure}[ht]
\begin{center}
\includegraphics[height=3.0in,width=3.2in]{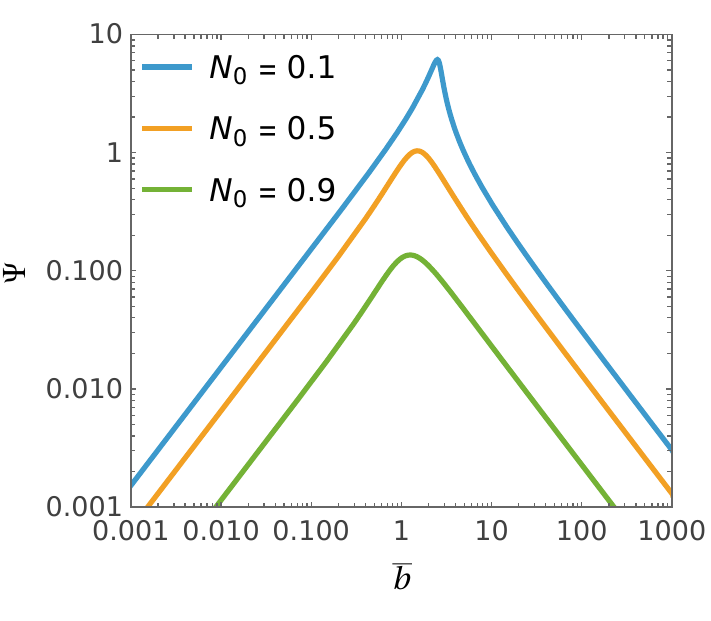}
\end{center}
\caption{Deflection angle $\Psi$ as a function 
of $\bar{b} = b/r_0$ 
for the exact solution defined by the metric components 
(\ref{fana}) and (\ref{hana}). 
Each line corresponds to 
$N_0 = 0.1$, $0.5$, and $0.9$, respectively.}
\label{anafig}
\end{figure}

In this Appendix, we compute the gravitational lensing deflection angle for the analytic model proposed in Refs.~\cite{DeFelice:2024seu,DeFelice:2025vef}. This model falls within the class of ESM theories described by the action (\ref{action}), but the coupling function $\mu(\phi)$ is specifically chosen to yield a 
solution for $N(r)=f(r)/h(r)$ of the form 
\be
N=\left( \frac{r^4 +\sqrt{N_0}\,r_0^4}
{r^4+r_0^4} \right)^2=
\left( \frac{x^4+\sqrt{N_0}}{x^4+1} \right)^2\,,
\label{Nr}
\ee
where $x=r/r_0$, and $N_0$ is constant in 
the range $0<N_0<1$.
In this case, the metric functions $f$ and $h$, as well as the radial derivatives of $\phi$ and $A_0$, 
admit exact solutions, which are given by
\ba
f &=& \frac{1}{8x^2}
\left[ 4\sqrt{2} \Delta _1
\bigl(\sqrt{N_0}-1\bigr)x
\right.\nonumber\\
&&+\left.\Delta_1^2 \bigl(\sqrt{N_0}-1\bigr)^2
+8 x^2 \right] 
\,,\label{fana}\\
h &=&\frac{\left(x^4+1\right)^2}
{8x^2 \left( x^4+\sqrt{N_0} 
\right)^2}
\left[ 4\sqrt{2} \Delta _1
\bigl(\sqrt{N_0}-1\bigr)x
\right.\nonumber\\
&&+\left.\Delta_1^2 \bigl(\sqrt{N_0}-1\bigr)^2
+8 x^2 \right] \,,
\label{hana}\\
\phi' &=&
\Mpl \sqrt{\frac{N'}{rN}}\,,
\label{rphi}\\
A_0' &=& -\frac{\Mpl^2 
[2(rh'+h-1)N+rhN']}{q_E \sqrt{N}}\,,
\label{rA02}
\ea
where 
\ba
\Delta_1 &\equiv& \arctan \left( \sqrt{2}x+1 \right)
+\arctan \left( \sqrt{2}x-1 \right) \nonumber \\
&&+{\rm arctanh} \left[  \sqrt{2}x/(x^2+1) \right]\,.
\ea
To realize this solution, the coupling $\mu(\phi)$ differs from that given in Eq.~(\ref{muform}). 
The leading-order behavior of $\mu(\phi)$ 
near $r=0$ is 
$\mu(\phi)=\bar{q}_E^2 \Mpl^3 \lambda_0
/[4(\phi-\phi_0)^3]$, where 
$\lambda_0=\sqrt{2(1-\sqrt{N_0})/\sqrt{N_0}}$, which corresponds to the case $m=2$ in Eq.~(\ref{mur=0}).
At spatial infinity, $\mu(\phi)$ approaches a constant value $\mu_{\infty}$.
The deviation from the coupling in Eq.~(\ref{muform}) arises at intermediate radii, where $\mu(\phi)$ attains a minimum in this analytic 
model \cite{DeFelice:2025vef}. 

In the above analytic model, the density profile of the ECOs exhibits a shell-like structure, with a peak 
around $r = r_0$. 
This suggests that the gravitational-lensing deflection angle $\Psi$ should attain a maximum for an impact parameter $b$ of order $r_0$.  
In Fig.~\ref{anafig}, we plot $\Psi$ as a function of $\bar{b} = b/r_0$ for three different values of $N_0$.  
Indeed, $\Psi$ exhibits peaks around $\bar{b} = \mathcal{O}(1)$, whose heights increase as $N_0$ decreases.  
For $N_0$ in the range $0.04163 < N_0 < 1$, no light rings appear. As $N_0$ approaches its minimum value, $0.04163$, the maximum of $\Psi$ can reach (and become larger than)
$\mathcal{O}(10)$. These properties are similar to those of the ECOs realized with the coupling~(\ref{muform}).

\bibliographystyle{mybibstyle}
\bibliography{bib}

\begin{thebibliography}{74}%
\makeatletter
\providecommand \@ifxundefined [1]{%
 \@ifx{#1\undefined}
}%
\providecommand \@ifnum [1]{%
 \ifnum #1\expandafter \@firstoftwo
 \else \expandafter \@secondoftwo
 \fi
}%
\providecommand \@ifx [1]{%
 \ifx #1\expandafter \@firstoftwo
 \else \expandafter \@secondoftwo
 \fi
}%
\providecommand \natexlab [1]{#1}%
\providecommand \enquote  [1]{``#1''}%
\providecommand \bibnamefont  [1]{#1}%
\providecommand \bibfnamefont [1]{#1}%
\providecommand \citenamefont [1]{#1}%
\providecommand \href@noop [0]{\@secondoftwo}%
\providecommand \href [0]{\begingroup \@sanitize@url \@href}%
\providecommand \@href[1]{\@@startlink{#1}\@@href}%
\providecommand \@@href[1]{\endgroup#1\@@endlink}%
\providecommand \@sanitize@url [0]{\catcode `\\12\catcode `\$12\catcode `\&12\catcode `\#12\catcode `\^12\catcode `\_12\catcode `\%12\relax}%
\providecommand \@@startlink[1]{}%
\providecommand \@@endlink[0]{}%
\providecommand \url  [0]{\begingroup\@sanitize@url \@url }%
\providecommand \@url [1]{\endgroup\@href {#1}{\urlprefix }}%
\providecommand \urlprefix  [0]{URL }%
\providecommand \Eprint [0]{\href }%
\providecommand \doibase [0]{http://dx.doi.org/}%
\providecommand \selectlanguage [0]{\@gobble}%
\providecommand \bibinfo  [0]{\@secondoftwo}%
\providecommand \bibfield  [0]{\@secondoftwo}%
\providecommand \translation [1]{[#1]}%
\providecommand \BibitemOpen [0]{}%
\providecommand \bibitemStop [0]{}%
\providecommand \bibitemNoStop [0]{.\EOS\space}%
\providecommand \EOS [0]{\spacefactor3000\relax}%
\providecommand \BibitemShut  [1]{\csname bibitem#1\endcsname}%
\let\auto@bib@innerbib\@empty
\bibitem [{\citenamefont {Will}(1993)}]{Will1993}%
  \BibitemOpen
  \bibfield  {author} {\bibinfo {author} {\bibfnamefont {C.~M.}\ \bibnamefont {Will}},\ }\href@noop {} {\emph {\bibinfo {title} {Theory and Experiment in Gravitational Physics}}},\ \bibinfo {edition} {2nd}\ ed.\ (\bibinfo  {publisher} {Cambridge University Press},\ \bibinfo {address} {Cambridge},\ \bibinfo {year} {1993})\BibitemShut {NoStop}%
\bibitem [{\citenamefont {Will}(2014)}]{Will:2014kxa}%
  \BibitemOpen
  \bibfield  {author} {\bibinfo {author} {\bibfnamefont {C.~M.}\ \bibnamefont {Will}},\ }\href {\doibase 10.12942/lrr-2014-4} {\bibfield  {journal} {\bibinfo  {journal} {\emph {Living Rev. Rel.}}\ }\textbf {\bibinfo {volume} {17}},\ \bibinfo {pages} {4} (\bibinfo {year} {2014})},\ \Eprint {http://arxiv.org/abs/1403.7377} {arXiv:1403.7377 [gr-qc]} \BibitemShut {NoStop}%
\bibitem [{\citenamefont {Abbott}\ \emph {et~al.}(2016)\citenamefont {Abbott} \emph {et~al.}}]{LIGOScientific:2016aoc}%
  \BibitemOpen
  \bibfield  {author} {\bibinfo {author} {\bibfnamefont {B.~P.}\ \bibnamefont {Abbott}} \emph {et~al.} (\bibinfo {collaboration} {LIGO Scientific, Virgo}),\ }\href {\doibase 10.1103/PhysRevLett.116.061102} {\bibfield  {journal} {\bibinfo  {journal} {\emph {Phys. Rev. Lett.}}\ }\textbf {\bibinfo {volume} {116}},\ \bibinfo {pages} {061102} (\bibinfo {year} {2016})},\ \Eprint {http://arxiv.org/abs/1602.03837} {arXiv:1602.03837 [gr-qc]} \BibitemShut {NoStop}%
\bibitem [{\citenamefont {Abbott}\ \emph {et~al.}(2017)\citenamefont {Abbott} \emph {et~al.}}]{LIGOScientific:2017vwq}%
  \BibitemOpen
  \bibfield  {author} {\bibinfo {author} {\bibfnamefont {B.~P.}\ \bibnamefont {Abbott}} \emph {et~al.} (\bibinfo {collaboration} {LIGO Scientific, Virgo}),\ }\href {\doibase 10.1103/PhysRevLett.119.161101} {\bibfield  {journal} {\bibinfo  {journal} {\emph {Phys. Rev. Lett.}}\ }\textbf {\bibinfo {volume} {119}},\ \bibinfo {pages} {161101} (\bibinfo {year} {2017})},\ \Eprint {http://arxiv.org/abs/1710.05832} {arXiv:1710.05832 [gr-qc]} \BibitemShut {NoStop}%
\bibitem [{\citenamefont {Blumenthal}\ \emph {et~al.}(1984)\citenamefont {Blumenthal}, \citenamefont {Faber}, \citenamefont {Primack},\ and\ \citenamefont {Rees}}]{Blumenthal:1984bp}%
  \BibitemOpen
  \bibfield  {author} {\bibinfo {author} {\bibfnamefont {G.~R.}\ \bibnamefont {Blumenthal}}, \bibinfo {author} {\bibfnamefont {S.~M.}\ \bibnamefont {Faber}}, \bibinfo {author} {\bibfnamefont {J.~R.}\ \bibnamefont {Primack}},  and \bibinfo {author} {\bibfnamefont {M.~J.}\ \bibnamefont {Rees}},\ }\href {\doibase 10.1038/311517a0} {\bibfield  {journal} {\bibinfo  {journal} {\emph {Nature}}\ }\textbf {\bibinfo {volume} {311}},\ \bibinfo {pages} {517} (\bibinfo {year} {1984})}\BibitemShut {NoStop}%
\bibitem [{\citenamefont {Jungman}\ \emph {et~al.}(1996)\citenamefont {Jungman}, \citenamefont {Kamionkowski},\ and\ \citenamefont {Griest}}]{Jungman:1995df}%
  \BibitemOpen
  \bibfield  {author} {\bibinfo {author} {\bibfnamefont {G.}~\bibnamefont {Jungman}}, \bibinfo {author} {\bibfnamefont {M.}~\bibnamefont {Kamionkowski}},  and \bibinfo {author} {\bibfnamefont {K.}~\bibnamefont {Griest}},\ }\href {\doibase 10.1016/0370-1573(95)00058-5} {\bibfield  {journal} {\bibinfo  {journal} {\emph {Phys. Rept.}}\ }\textbf {\bibinfo {volume} {267}},\ \bibinfo {pages} {195} (\bibinfo {year} {1996})},\ \Eprint {http://arxiv.org/abs/hep-ph/9506380} {arXiv:hep-ph/9506380} \BibitemShut {NoStop}%
\bibitem [{\citenamefont {Bertone}\ \emph {et~al.}(2005)\citenamefont {Bertone}, \citenamefont {Hooper},\ and\ \citenamefont {Silk}}]{Bertone:2004pz}%
  \BibitemOpen
  \bibfield  {author} {\bibinfo {author} {\bibfnamefont {G.}~\bibnamefont {Bertone}}, \bibinfo {author} {\bibfnamefont {D.}~\bibnamefont {Hooper}},  and \bibinfo {author} {\bibfnamefont {J.}~\bibnamefont {Silk}},\ }\href {\doibase 10.1016/j.physrep.2004.08.031} {\bibfield  {journal} {\bibinfo  {journal} {\emph {Phys. Rept.}}\ }\textbf {\bibinfo {volume} {405}},\ \bibinfo {pages} {279} (\bibinfo {year} {2005})},\ \Eprint {http://arxiv.org/abs/hep-ph/0404175} {arXiv:hep-ph/0404175} \BibitemShut {NoStop}%
\bibitem [{\citenamefont {Zwicky}(1937)}]{Zwicky:1937zza}%
  \BibitemOpen
  \bibfield  {author} {\bibinfo {author} {\bibfnamefont {F.}~\bibnamefont {Zwicky}},\ }\href {\doibase 10.1086/143864} {\bibfield  {journal} {\bibinfo  {journal} {\emph {Astrophys. J.}}\ }\textbf {\bibinfo {volume} {86}},\ \bibinfo {pages} {217} (\bibinfo {year} {1937})}\BibitemShut {NoStop}%
\bibitem [{\citenamefont {Rubin}\ and\ \citenamefont {Ford}(1970)}]{Rubin:1970zza}%
  \BibitemOpen
  \bibfield  {author} {\bibinfo {author} {\bibfnamefont {V.~C.}\ \bibnamefont {Rubin}} and \bibinfo {author} {\bibfnamefont {W.~K.}\ \bibnamefont {Ford}, \bibfnamefont {Jr.}},\ }\href {\doibase 10.1086/150317} {\bibfield  {journal} {\bibinfo  {journal} {\emph {Astrophys. J.}}\ }\textbf {\bibinfo {volume} {159}},\ \bibinfo {pages} {379} (\bibinfo {year} {1970})}\BibitemShut {NoStop}%
\bibitem [{\citenamefont {Sofue}\ and\ \citenamefont {Rubin}(2001)}]{Sofue:2000jx}%
  \BibitemOpen
  \bibfield  {author} {\bibinfo {author} {\bibfnamefont {Y.}~\bibnamefont {Sofue}} and \bibinfo {author} {\bibfnamefont {V.}~\bibnamefont {Rubin}},\ }\href {\doibase 10.1146/annurev.astro.39.1.137} {\bibfield  {journal} {\bibinfo  {journal} {\emph {Ann. Rev. Astron. Astrophys.}}\ }\textbf {\bibinfo {volume} {39}},\ \bibinfo {pages} {137} (\bibinfo {year} {2001})},\ \Eprint {http://arxiv.org/abs/astro-ph/0010594} {arXiv:astro-ph/0010594} \BibitemShut {NoStop}%
\bibitem [{\citenamefont {Bartelmann}\ and\ \citenamefont {Schneider}(2001)}]{Bartelmann:1999yn}%
  \BibitemOpen
  \bibfield  {author} {\bibinfo {author} {\bibfnamefont {M.}~\bibnamefont {Bartelmann}} and \bibinfo {author} {\bibfnamefont {P.}~\bibnamefont {Schneider}},\ }\href {\doibase 10.1016/S0370-1573(00)00082-X} {\bibfield  {journal} {\bibinfo  {journal} {\emph {Phys. Rept.}}\ }\textbf {\bibinfo {volume} {340}},\ \bibinfo {pages} {291} (\bibinfo {year} {2001})},\ \Eprint {http://arxiv.org/abs/astro-ph/9912508} {arXiv:astro-ph/9912508} \BibitemShut {NoStop}%
\bibitem [{\citenamefont {Clowe}\ \emph {et~al.}(2006)\citenamefont {Clowe}, \citenamefont {Bradac}, \citenamefont {Gonzalez}, \citenamefont {Markevitch}, \citenamefont {Randall}, \citenamefont {Jones},\ and\ \citenamefont {Zaritsky}}]{Clowe:2006eq}%
  \BibitemOpen
  \bibfield  {author} {\bibinfo {author} {\bibfnamefont {D.}~\bibnamefont {Clowe}}, \bibinfo {author} {\bibfnamefont {M.}~\bibnamefont {Bradac}}, \bibinfo {author} {\bibfnamefont {A.~H.}\ \bibnamefont {Gonzalez}}, \bibinfo {author} {\bibfnamefont {M.}~\bibnamefont {Markevitch}}, \bibinfo {author} {\bibfnamefont {S.~W.}\ \bibnamefont {Randall}}, \bibinfo {author} {\bibfnamefont {C.}~\bibnamefont {Jones}},  and \bibinfo {author} {\bibfnamefont {D.}~\bibnamefont {Zaritsky}},\ }\href {\doibase 10.1086/508162} {\bibfield  {journal} {\bibinfo  {journal} {\emph {Astrophys. J. Lett.}}\ }\textbf {\bibinfo {volume} {648}},\ \bibinfo {pages} {L109} (\bibinfo {year} {2006})},\ \Eprint {http://arxiv.org/abs/astro-ph/0608407} {arXiv:astro-ph/0608407} \BibitemShut {NoStop}%
\bibitem [{\citenamefont {Anderson}(2014)}]{Anderson2014}%
  \BibitemOpen
  \bibfield  {author} {\bibinfo {author} {\bibfnamefont {L.~e.~a.}\ \bibnamefont {Anderson}},\ }\href {\doibase 10.1093/mnras/stu523} {\bibfield  {journal} {\bibinfo  {journal} {\emph {Monthly Notices of the Royal Astronomical Society}}\ }\textbf {\bibinfo {volume} {441}},\ \bibinfo {pages} {24} (\bibinfo {year} {2014})}\BibitemShut {NoStop}%
\bibitem [{\citenamefont {Alam}(2021)}]{eBOSS2021}%
  \BibitemOpen
  \bibfield  {author} {\bibinfo {author} {\bibfnamefont {S.~e.~a.}\ \bibnamefont {Alam}},\ }\href {\doibase 10.1103/PhysRevD.103.083533} {\bibfield  {journal} {\bibinfo  {journal} {\emph {Physical Review D}}\ }\textbf {\bibinfo {volume} {103}},\ \bibinfo {pages} {083533} (\bibinfo {year} {2021})}\BibitemShut {NoStop}%
\bibitem [{\citenamefont {Abbott}\ \emph {et~al.}(2024)\citenamefont {Abbott} \emph {et~al.}}]{DES:2024jxu}%
  \BibitemOpen
  \bibfield  {author} {\bibinfo {author} {\bibfnamefont {T.~M.~C.}\ \bibnamefont {Abbott}} \emph {et~al.} (\bibinfo {collaboration} {DES}),\ }\href {\doibase 10.3847/2041-8213/ad6f9f} {\bibfield  {journal} {\bibinfo  {journal} {\emph {Astrophys. J. Lett.}}\ }\textbf {\bibinfo {volume} {973}},\ \bibinfo {pages} {L14} (\bibinfo {year} {2024})},\ \Eprint {http://arxiv.org/abs/2401.02929} {arXiv:2401.02929 [astro-ph.CO]} \BibitemShut {NoStop}%
\bibitem [{\citenamefont {Abdul~Karim}\ \emph {et~al.}(2025)\citenamefont {Abdul~Karim} \emph {et~al.}}]{DESI:2025zgx}%
  \BibitemOpen
  \bibfield  {author} {\bibinfo {author} {\bibfnamefont {M.}~\bibnamefont {Abdul~Karim}} \emph {et~al.} (\bibinfo {collaboration} {DESI}),\ }\href {\doibase 10.1103/tr6y-kpc6} {\bibfield  {journal} {\bibinfo  {journal} {\emph {Phys. Rev. D}}\ }\textbf {\bibinfo {volume} {112}},\ \bibinfo {pages} {083515} (\bibinfo {year} {2025})},\ \Eprint {http://arxiv.org/abs/2503.14738} {arXiv:2503.14738 [astro-ph.CO]} \BibitemShut {NoStop}%
\bibitem [{\citenamefont {Hinshaw}\ \emph {et~al.}(2013)\citenamefont {Hinshaw} \emph {et~al.}}]{WMAP:2012nax}%
  \BibitemOpen
  \bibfield  {author} {\bibinfo {author} {\bibfnamefont {G.}~\bibnamefont {Hinshaw}} \emph {et~al.} (\bibinfo {collaboration} {WMAP}),\ }\href {\doibase 10.1088/0067-0049/208/2/19} {\bibfield  {journal} {\bibinfo  {journal} {\emph {Astrophys. J. Suppl.}}\ }\textbf {\bibinfo {volume} {208}},\ \bibinfo {pages} {19} (\bibinfo {year} {2013})},\ \Eprint {http://arxiv.org/abs/1212.5226} {arXiv:1212.5226 [astro-ph.CO]} \BibitemShut {NoStop}%
\bibitem [{\citenamefont {Aghanim}\ \emph {et~al.}(2020)\citenamefont {Aghanim} \emph {et~al.}}]{Planck:2018vyg}%
  \BibitemOpen
  \bibfield  {author} {\bibinfo {author} {\bibfnamefont {N.}~\bibnamefont {Aghanim}} \emph {et~al.} (\bibinfo {collaboration} {Planck}),\ }\href {\doibase 10.1051/0004-6361/201833910} {\bibfield  {journal} {\bibinfo  {journal} {\emph {Astron. Astrophys.}}\ }\textbf {\bibinfo {volume} {641}},\ \bibinfo {pages} {A6} (\bibinfo {year} {2020})},\ \bibinfo {note} {[Erratum: Astron.Astrophys. 652, C4 (2021)]},\ \Eprint {http://arxiv.org/abs/1807.06209} {arXiv:1807.06209 [astro-ph.CO]} \BibitemShut {NoStop}%
\bibitem [{\citenamefont {Yoneya}(1974)}]{Yoneya:1974jg}%
  \BibitemOpen
  \bibfield  {author} {\bibinfo {author} {\bibfnamefont {T.}~\bibnamefont {Yoneya}},\ }\href {\doibase 10.1143/PTP.51.1907} {\bibfield  {journal} {\bibinfo  {journal} {\emph {Prog. Theor. Phys.}}\ }\textbf {\bibinfo {volume} {51}},\ \bibinfo {pages} {1907} (\bibinfo {year} {1974})}\BibitemShut {NoStop}%
\bibitem [{\citenamefont {Scherk}\ and\ \citenamefont {Schwarz}(1974)}]{Scherk:1974ca}%
  \BibitemOpen
  \bibfield  {author} {\bibinfo {author} {\bibfnamefont {J.}~\bibnamefont {Scherk}} and \bibinfo {author} {\bibfnamefont {J.~H.}\ \bibnamefont {Schwarz}},\ }\href {\doibase 10.1016/0550-3213(74)90010-8} {\bibfield  {journal} {\bibinfo  {journal} {\emph {Nucl. Phys. B}}\ }\textbf {\bibinfo {volume} {81}},\ \bibinfo {pages} {118} (\bibinfo {year} {1974})}\BibitemShut {NoStop}%
\bibitem [{\citenamefont {Callan}\ \emph {et~al.}(1987)\citenamefont {Callan}, \citenamefont {Lovelace}, \citenamefont {Nappi},\ and\ \citenamefont {Yost}}]{Callan:1986bc}%
  \BibitemOpen
  \bibfield  {author} {\bibinfo {author} {\bibfnamefont {C.~G.}\ \bibnamefont {Callan}, \bibfnamefont {Jr.}}, \bibinfo {author} {\bibfnamefont {C.}~\bibnamefont {Lovelace}}, \bibinfo {author} {\bibfnamefont {C.~R.}\ \bibnamefont {Nappi}},  and \bibinfo {author} {\bibfnamefont {S.~A.}\ \bibnamefont {Yost}},\ }\href {\doibase 10.1016/0550-3213(87)90227-6} {\bibfield  {journal} {\bibinfo  {journal} {\emph {Nucl. Phys. B}}\ }\textbf {\bibinfo {volume} {288}},\ \bibinfo {pages} {525} (\bibinfo {year} {1987})}\BibitemShut {NoStop}%
\bibitem [{\citenamefont {Gaillard}\ and\ \citenamefont {Zumino}(1981)}]{Gaillard:1981rj}%
  \BibitemOpen
  \bibfield  {author} {\bibinfo {author} {\bibfnamefont {M.~K.}\ \bibnamefont {Gaillard}} and \bibinfo {author} {\bibfnamefont {B.}~\bibnamefont {Zumino}},\ }\href {\doibase 10.1016/0550-3213(81)90527-7} {\bibfield  {journal} {\bibinfo  {journal} {\emph {Nucl. Phys. B}}\ }\textbf {\bibinfo {volume} {193}},\ \bibinfo {pages} {221} (\bibinfo {year} {1981})}\BibitemShut {NoStop}%
\bibitem [{\citenamefont {Duff}\ \emph {et~al.}(1996)\citenamefont {Duff}, \citenamefont {Liu},\ and\ \citenamefont {Rahmfeld}}]{Duff:1995sm}%
  \BibitemOpen
  \bibfield  {author} {\bibinfo {author} {\bibfnamefont {M.~J.}\ \bibnamefont {Duff}}, \bibinfo {author} {\bibfnamefont {J.~T.}\ \bibnamefont {Liu}},  and \bibinfo {author} {\bibfnamefont {J.}~\bibnamefont {Rahmfeld}},\ }\href {\doibase 10.1016/0550-3213(95)00555-2} {\bibfield  {journal} {\bibinfo  {journal} {\emph {Nucl. Phys. B}}\ }\textbf {\bibinfo {volume} {459}},\ \bibinfo {pages} {125} (\bibinfo {year} {1996})},\ \Eprint {http://arxiv.org/abs/hep-th/9508094} {arXiv:hep-th/9508094} \BibitemShut {NoStop}%
\bibitem [{\citenamefont {Andrianopoli}\ \emph {et~al.}(1997)\citenamefont {Andrianopoli}, \citenamefont {Bertolini}, \citenamefont {Ceresole}, \citenamefont {D'Auria}, \citenamefont {Ferrara}, \citenamefont {Fre},\ and\ \citenamefont {Magri}}]{Andrianopoli:1996cm}%
  \BibitemOpen
  \bibfield  {author} {\bibinfo {author} {\bibfnamefont {L.}~\bibnamefont {Andrianopoli}}, \bibinfo {author} {\bibfnamefont {M.}~\bibnamefont {Bertolini}}, \bibinfo {author} {\bibfnamefont {A.}~\bibnamefont {Ceresole}}, \bibinfo {author} {\bibfnamefont {R.}~\bibnamefont {D'Auria}}, \bibinfo {author} {\bibfnamefont {S.}~\bibnamefont {Ferrara}}, \bibinfo {author} {\bibfnamefont {P.}~\bibnamefont {Fre}},  and \bibinfo {author} {\bibfnamefont {T.}~\bibnamefont {Magri}},\ }\href {\doibase 10.1016/S0393-0440(97)00002-8} {\bibfield  {journal} {\bibinfo  {journal} {\emph {J. Geom. Phys.}}\ }\textbf {\bibinfo {volume} {23}},\ \bibinfo {pages} {111} (\bibinfo {year} {1997})},\ \Eprint {http://arxiv.org/abs/hep-th/9605032} {arXiv:hep-th/9605032} \BibitemShut {NoStop}%
\bibitem [{\citenamefont {Gibbons}\ and\ \citenamefont {Maeda}(1988)}]{Gibbons:1987ps}%
  \BibitemOpen
  \bibfield  {author} {\bibinfo {author} {\bibfnamefont {G.~W.}\ \bibnamefont {Gibbons}} and \bibinfo {author} {\bibfnamefont {K.-i.}\ \bibnamefont {Maeda}},\ }\href {\doibase 10.1016/0550-3213(88)90006-5} {\bibfield  {journal} {\bibinfo  {journal} {\emph {Nucl. Phys. B}}\ }\textbf {\bibinfo {volume} {298}},\ \bibinfo {pages} {741} (\bibinfo {year} {1988})}\BibitemShut {NoStop}%
\bibitem [{\citenamefont {Garfinkle}\ \emph {et~al.}(1991)\citenamefont {Garfinkle}, \citenamefont {Horowitz},\ and\ \citenamefont {Strominger}}]{Garfinkle:1990qj}%
  \BibitemOpen
  \bibfield  {author} {\bibinfo {author} {\bibfnamefont {D.}~\bibnamefont {Garfinkle}}, \bibinfo {author} {\bibfnamefont {G.~T.}\ \bibnamefont {Horowitz}},  and \bibinfo {author} {\bibfnamefont {A.}~\bibnamefont {Strominger}},\ }\href {\doibase 10.1103/PhysRevD.43.3140} {\bibfield  {journal} {\bibinfo  {journal} {\emph {Phys. Rev. D}}\ }\textbf {\bibinfo {volume} {43}},\ \bibinfo {pages} {3140} (\bibinfo {year} {1991})},\ \bibinfo {note} {[Erratum: Phys.Rev.D 45, 3888 (1992)]}\BibitemShut {NoStop}%
\bibitem [{\citenamefont {De~Felice}\ and\ \citenamefont {Tsujikawa}(2025{\natexlab{a}})}]{DeFelice:2024ops}%
  \BibitemOpen
  \bibfield  {author} {\bibinfo {author} {\bibfnamefont {A.}~\bibnamefont {De~Felice}} and \bibinfo {author} {\bibfnamefont {S.}~\bibnamefont {Tsujikawa}},\ }\href {\doibase 10.1103/PhysRevD.111.064051} {\bibfield  {journal} {\bibinfo  {journal} {\emph {Phys. Rev. D}}\ }\textbf {\bibinfo {volume} {111}},\ \bibinfo {pages} {064051} (\bibinfo {year} {2025}{\natexlab{a}})},\ \Eprint {http://arxiv.org/abs/2412.04754} {arXiv:2412.04754 [gr-qc]} \BibitemShut {NoStop}%
\bibitem [{\citenamefont {De~Felice}\ and\ \citenamefont {Tsujikawa}(2025{\natexlab{b}})}]{DeFelice:2024seu}%
  \BibitemOpen
  \bibfield  {author} {\bibinfo {author} {\bibfnamefont {A.}~\bibnamefont {De~Felice}} and \bibinfo {author} {\bibfnamefont {S.}~\bibnamefont {Tsujikawa}},\ }\href {\doibase 10.1103/PhysRevLett.134.081401} {\bibfield  {journal} {\bibinfo  {journal} {\emph {Phys. Rev. Lett.}}\ }\textbf {\bibinfo {volume} {134}},\ \bibinfo {pages} {081401} (\bibinfo {year} {2025}{\natexlab{b}})},\ \Eprint {http://arxiv.org/abs/2410.00314} {arXiv:2410.00314 [gr-qc]} \BibitemShut {NoStop}%
\bibitem [{\citenamefont {Schunck}\ and\ \citenamefont {Mielke}(2003)}]{Schunck:2003kk}%
  \BibitemOpen
  \bibfield  {author} {\bibinfo {author} {\bibfnamefont {F.~E.}\ \bibnamefont {Schunck}} and \bibinfo {author} {\bibfnamefont {E.~W.}\ \bibnamefont {Mielke}},\ }\href {\doibase 10.1088/0264-9381/20/20/201} {\bibfield  {journal} {\bibinfo  {journal} {\emph {Class. Quant. Grav.}}\ }\textbf {\bibinfo {volume} {20}},\ \bibinfo {pages} {R301} (\bibinfo {year} {2003})},\ \Eprint {http://arxiv.org/abs/0801.0307} {arXiv:0801.0307 [astro-ph]} \BibitemShut {NoStop}%
\bibitem [{\citenamefont {Liebling}\ and\ \citenamefont {Palenzuela}(2023)}]{Liebling:2012fv}%
  \BibitemOpen
  \bibfield  {author} {\bibinfo {author} {\bibfnamefont {S.~L.}\ \bibnamefont {Liebling}} and \bibinfo {author} {\bibfnamefont {C.}~\bibnamefont {Palenzuela}},\ }\href {\doibase 10.1007/s41114-023-00043-4} {\bibfield  {journal} {\bibinfo  {journal} {\emph {Living Rev. Rel.}}\ }\textbf {\bibinfo {volume} {26}},\ \bibinfo {pages} {1} (\bibinfo {year} {2023})},\ \Eprint {http://arxiv.org/abs/1202.5809} {arXiv:1202.5809 [gr-qc]} \BibitemShut {NoStop}%
\bibitem [{\citenamefont {Barack}\ \emph {et~al.}(2019)\citenamefont {Barack} \emph {et~al.}}]{Barack:2018yly}%
  \BibitemOpen
  \bibfield  {author} {\bibinfo {author} {\bibfnamefont {L.}~\bibnamefont {Barack}} \emph {et~al.},\ }\href {\doibase 10.1088/1361-6382/ab0587} {\bibfield  {journal} {\bibinfo  {journal} {\emph {Class. Quant. Grav.}}\ }\textbf {\bibinfo {volume} {36}},\ \bibinfo {pages} {143001} (\bibinfo {year} {2019})},\ \Eprint {http://arxiv.org/abs/1806.05195} {arXiv:1806.05195 [gr-qc]} \BibitemShut {NoStop}%
\bibitem [{\citenamefont {Cardoso}\ and\ \citenamefont {Pani}(2019)}]{Cardoso:2019rvt}%
  \BibitemOpen
  \bibfield  {author} {\bibinfo {author} {\bibfnamefont {V.}~\bibnamefont {Cardoso}} and \bibinfo {author} {\bibfnamefont {P.}~\bibnamefont {Pani}},\ }\href {\doibase 10.1007/s41114-019-0020-4} {\bibfield  {journal} {\bibinfo  {journal} {\emph {Living Rev. Rel.}}\ }\textbf {\bibinfo {volume} {22}},\ \bibinfo {pages} {4} (\bibinfo {year} {2019})},\ \Eprint {http://arxiv.org/abs/1904.05363} {arXiv:1904.05363 [gr-qc]} \BibitemShut {NoStop}%
\bibitem [{\citenamefont {Derrick}(1964)}]{Derrick:1964ww}%
  \BibitemOpen
  \bibfield  {author} {\bibinfo {author} {\bibfnamefont {G.~H.}\ \bibnamefont {Derrick}},\ }\href {\doibase 10.1063/1.1704233} {\bibfield  {journal} {\bibinfo  {journal} {\emph {J. Math. Phys.}}\ }\textbf {\bibinfo {volume} {5}},\ \bibinfo {pages} {1252} (\bibinfo {year} {1964})}\BibitemShut {NoStop}%
\bibitem [{\citenamefont {Diez-Tejedor}\ and\ \citenamefont {Gonzalez-Morales}(2013)}]{Diez-Tejedor:2013sza}%
  \BibitemOpen
  \bibfield  {author} {\bibinfo {author} {\bibfnamefont {A.}~\bibnamefont {Diez-Tejedor}} and \bibinfo {author} {\bibfnamefont {A.~X.}\ \bibnamefont {Gonzalez-Morales}},\ }\href {\doibase 10.1103/PhysRevD.88.067302} {\bibfield  {journal} {\bibinfo  {journal} {\emph {Phys. Rev. D}}\ }\textbf {\bibinfo {volume} {88}},\ \bibinfo {pages} {067302} (\bibinfo {year} {2013})},\ \Eprint {http://arxiv.org/abs/1306.4400} {arXiv:1306.4400 [gr-qc]} \BibitemShut {NoStop}%
\bibitem [{\citenamefont {Wheeler}(1955)}]{Wheeler:1955zz}%
  \BibitemOpen
  \bibfield  {author} {\bibinfo {author} {\bibfnamefont {J.~A.}\ \bibnamefont {Wheeler}},\ }\href {\doibase 10.1103/PhysRev.97.511} {\bibfield  {journal} {\bibinfo  {journal} {\emph {Phys. Rev.}}\ }\textbf {\bibinfo {volume} {97}},\ \bibinfo {pages} {511} (\bibinfo {year} {1955})}\BibitemShut {NoStop}%
\bibitem [{\citenamefont {Power}\ and\ \citenamefont {Wheeler}(1957)}]{Power:1957zz}%
  \BibitemOpen
  \bibfield  {author} {\bibinfo {author} {\bibfnamefont {E.~A.}\ \bibnamefont {Power}} and \bibinfo {author} {\bibfnamefont {J.~A.}\ \bibnamefont {Wheeler}},\ }\href {\doibase 10.1103/RevModPhys.29.480} {\bibfield  {journal} {\bibinfo  {journal} {\emph {Rev. Mod. Phys.}}\ }\textbf {\bibinfo {volume} {29}},\ \bibinfo {pages} {480} (\bibinfo {year} {1957})}\BibitemShut {NoStop}%
\bibitem [{\citenamefont {Kaup}(1968)}]{Kaup:1968zz}%
  \BibitemOpen
  \bibfield  {author} {\bibinfo {author} {\bibfnamefont {D.~J.}\ \bibnamefont {Kaup}},\ }\href {\doibase 10.1103/PhysRev.172.1331} {\bibfield  {journal} {\bibinfo  {journal} {\emph {Phys. Rev.}}\ }\textbf {\bibinfo {volume} {172}},\ \bibinfo {pages} {1331} (\bibinfo {year} {1968})}\BibitemShut {NoStop}%
\bibitem [{\citenamefont {Ruffini}\ and\ \citenamefont {Bonazzola}(1969)}]{Ruffini:1969qy}%
  \BibitemOpen
  \bibfield  {author} {\bibinfo {author} {\bibfnamefont {R.}~\bibnamefont {Ruffini}} and \bibinfo {author} {\bibfnamefont {S.}~\bibnamefont {Bonazzola}},\ }\href {\doibase 10.1103/PhysRev.187.1767} {\bibfield  {journal} {\bibinfo  {journal} {\emph {Phys. Rev.}}\ }\textbf {\bibinfo {volume} {187}},\ \bibinfo {pages} {1767} (\bibinfo {year} {1969})}\BibitemShut {NoStop}%
\bibitem [{\citenamefont {Brito}\ \emph {et~al.}(2016)\citenamefont {Brito}, \citenamefont {Cardoso}, \citenamefont {Herdeiro},\ and\ \citenamefont {Radu}}]{Brito:2015pxa}%
  \BibitemOpen
  \bibfield  {author} {\bibinfo {author} {\bibfnamefont {R.}~\bibnamefont {Brito}}, \bibinfo {author} {\bibfnamefont {V.}~\bibnamefont {Cardoso}}, \bibinfo {author} {\bibfnamefont {C.~A.~R.}\ \bibnamefont {Herdeiro}},  and \bibinfo {author} {\bibfnamefont {E.}~\bibnamefont {Radu}},\ }\href {\doibase 10.1016/j.physletb.2015.11.051} {\bibfield  {journal} {\bibinfo  {journal} {\emph {Phys. Lett. B}}\ }\textbf {\bibinfo {volume} {752}},\ \bibinfo {pages} {291} (\bibinfo {year} {2016})},\ \Eprint {http://arxiv.org/abs/1508.05395} {arXiv:1508.05395 [gr-qc]} \BibitemShut {NoStop}%
\bibitem [{\citenamefont {Herdeiro}\ \emph {et~al.}(2016)\citenamefont {Herdeiro}, \citenamefont {Radu},\ and\ \citenamefont {R{\'u}narsson}}]{Herdeiro:2016tmi}%
  \BibitemOpen
  \bibfield  {author} {\bibinfo {author} {\bibfnamefont {C.}~\bibnamefont {Herdeiro}}, \bibinfo {author} {\bibfnamefont {E.}~\bibnamefont {Radu}},  and \bibinfo {author} {\bibfnamefont {H.}~\bibnamefont {R{\'u}narsson}},\ }\href {\doibase 10.1088/0264-9381/33/15/154001} {\bibfield  {journal} {\bibinfo  {journal} {\emph {Class. Quant. Grav.}}\ }\textbf {\bibinfo {volume} {33}},\ \bibinfo {pages} {154001} (\bibinfo {year} {2016})},\ \Eprint {http://arxiv.org/abs/1603.02687} {arXiv:1603.02687 [gr-qc]} \BibitemShut {NoStop}%
\bibitem [{\citenamefont {Herdeiro}\ and\ \citenamefont {Oliveira}(2019)}]{Herdeiro:2019oqp}%
  \BibitemOpen
  \bibfield  {author} {\bibinfo {author} {\bibfnamefont {C.~A.~R.}\ \bibnamefont {Herdeiro}} and \bibinfo {author} {\bibfnamefont {J.~M.~S.}\ \bibnamefont {Oliveira}},\ }\href {\doibase 10.1088/1361-6382/ab1859} {\bibfield  {journal} {\bibinfo  {journal} {\emph {Class. Quant. Grav.}}\ }\textbf {\bibinfo {volume} {36}},\ \bibinfo {pages} {105015} (\bibinfo {year} {2019})},\ \Eprint {http://arxiv.org/abs/1902.07721} {arXiv:1902.07721 [gr-qc]} \BibitemShut {NoStop}%
\bibitem [{\citenamefont {Herdeiro}\ \emph {et~al.}(2020)\citenamefont {Herdeiro}, \citenamefont {Oliveira},\ and\ \citenamefont {Radu}}]{Herdeiro:2019iwl}%
  \BibitemOpen
  \bibfield  {author} {\bibinfo {author} {\bibfnamefont {C.~A.~R.}\ \bibnamefont {Herdeiro}}, \bibinfo {author} {\bibfnamefont {J.~M.~S.}\ \bibnamefont {Oliveira}},  and \bibinfo {author} {\bibfnamefont {E.}~\bibnamefont {Radu}},\ }\href {\doibase 10.1140/epjc/s10052-019-7583-9} {\bibfield  {journal} {\bibinfo  {journal} {\emph {Eur. Phys. J. C}}\ }\textbf {\bibinfo {volume} {80}},\ \bibinfo {pages} {23} (\bibinfo {year} {2020})},\ \Eprint {http://arxiv.org/abs/1910.11021} {arXiv:1910.11021 [gr-qc]} \BibitemShut {NoStop}%
\bibitem [{\citenamefont {De~Felice}\ and\ \citenamefont {Tsujikawa}(2026)}]{DeFelice:2025vef}%
  \BibitemOpen
  \bibfield  {author} {\bibinfo {author} {\bibfnamefont {A.}~\bibnamefont {De~Felice}} and \bibinfo {author} {\bibfnamefont {S.}~\bibnamefont {Tsujikawa}},\ }\href {\doibase 10.1103/7tmb-kp11} {\bibfield  {journal} {\bibinfo  {journal} {\emph {Phys. Rev. D}}\ }\textbf {\bibinfo {volume} {113}},\ \bibinfo {pages} {024014} (\bibinfo {year} {2026})},\ \Eprint {http://arxiv.org/abs/2511.14207} {arXiv:2511.14207 [gr-qc]} \BibitemShut {NoStop}%
\bibitem [{\citenamefont {Gralla}\ \emph {et~al.}(2019)\citenamefont {Gralla}, \citenamefont {Holz},\ and\ \citenamefont {Wald}}]{Gralla:2019xty}%
  \BibitemOpen
  \bibfield  {author} {\bibinfo {author} {\bibfnamefont {S.~E.}\ \bibnamefont {Gralla}}, \bibinfo {author} {\bibfnamefont {D.~E.}\ \bibnamefont {Holz}},  and \bibinfo {author} {\bibfnamefont {R.~M.}\ \bibnamefont {Wald}},\ }\href {\doibase 10.1103/PhysRevD.100.024018} {\bibfield  {journal} {\bibinfo  {journal} {\emph {Phys. Rev. D}}\ }\textbf {\bibinfo {volume} {100}},\ \bibinfo {pages} {024018} (\bibinfo {year} {2019})},\ \Eprint {http://arxiv.org/abs/1906.00873} {arXiv:1906.00873 [astro-ph.HE]} \BibitemShut {NoStop}%
\bibitem [{\citenamefont {Gralla}\ \emph {et~al.}(2020)\citenamefont {Gralla}, \citenamefont {Lupsasca},\ and\ \citenamefont {Marrone}}]{Gralla:2020srx}%
  \BibitemOpen
  \bibfield  {author} {\bibinfo {author} {\bibfnamefont {S.~E.}\ \bibnamefont {Gralla}}, \bibinfo {author} {\bibfnamefont {A.}~\bibnamefont {Lupsasca}},  and \bibinfo {author} {\bibfnamefont {D.~P.}\ \bibnamefont {Marrone}},\ }\href {\doibase 10.1103/PhysRevD.102.124004} {\bibfield  {journal} {\bibinfo  {journal} {\emph {Phys. Rev. D}}\ }\textbf {\bibinfo {volume} {102}},\ \bibinfo {pages} {124004} (\bibinfo {year} {2020})},\ \Eprint {http://arxiv.org/abs/2008.03879} {arXiv:2008.03879 [gr-qc]} \BibitemShut {NoStop}%
\bibitem [{\citenamefont {Gan}\ \emph {et~al.}(2021)\citenamefont {Gan}, \citenamefont {Wang}, \citenamefont {Wu},\ and\ \citenamefont {Yang}}]{Gan:2021xdl}%
  \BibitemOpen
  \bibfield  {author} {\bibinfo {author} {\bibfnamefont {Q.}~\bibnamefont {Gan}}, \bibinfo {author} {\bibfnamefont {P.}~\bibnamefont {Wang}}, \bibinfo {author} {\bibfnamefont {H.}~\bibnamefont {Wu}},  and \bibinfo {author} {\bibfnamefont {H.}~\bibnamefont {Yang}},\ }\href {\doibase 10.1103/PhysRevD.104.044049} {\bibfield  {journal} {\bibinfo  {journal} {\emph {Phys. Rev. D}}\ }\textbf {\bibinfo {volume} {104}},\ \bibinfo {pages} {044049} (\bibinfo {year} {2021})},\ \Eprint {http://arxiv.org/abs/2105.11770} {arXiv:2105.11770 [gr-qc]} \BibitemShut {NoStop}%
\bibitem [{\citenamefont {Paugnat}\ \emph {et~al.}(2022)\citenamefont {Paugnat}, \citenamefont {Lupsasca}, \citenamefont {Vincent},\ and\ \citenamefont {Wielgus}}]{Paugnat:2022qzy}%
  \BibitemOpen
  \bibfield  {author} {\bibinfo {author} {\bibfnamefont {H.}~\bibnamefont {Paugnat}}, \bibinfo {author} {\bibfnamefont {A.}~\bibnamefont {Lupsasca}}, \bibinfo {author} {\bibfnamefont {F.}~\bibnamefont {Vincent}},  and \bibinfo {author} {\bibfnamefont {M.}~\bibnamefont {Wielgus}},\ }\href {\doibase 10.1051/0004-6361/202244216} {\bibfield  {journal} {\bibinfo  {journal} {\emph {Astron. Astrophys.}}\ }\textbf {\bibinfo {volume} {668}},\ \bibinfo {pages} {A11} (\bibinfo {year} {2022})},\ \Eprint {http://arxiv.org/abs/2206.02781} {arXiv:2206.02781 [astro-ph.HE]} \BibitemShut {NoStop}%
\bibitem [{\citenamefont {Staelens}\ \emph {et~al.}(2023)\citenamefont {Staelens}, \citenamefont {Mayerson}, \citenamefont {Bacchini}, \citenamefont {Ripperda},\ and\ \citenamefont {K{\"u}chler}}]{Staelens:2023jgr}%
  \BibitemOpen
  \bibfield  {author} {\bibinfo {author} {\bibfnamefont {S.}~\bibnamefont {Staelens}}, \bibinfo {author} {\bibfnamefont {D.~R.}\ \bibnamefont {Mayerson}}, \bibinfo {author} {\bibfnamefont {F.}~\bibnamefont {Bacchini}}, \bibinfo {author} {\bibfnamefont {B.}~\bibnamefont {Ripperda}},  and \bibinfo {author} {\bibfnamefont {L.}~\bibnamefont {K{\"u}chler}},\ }\href {\doibase 10.1103/PhysRevD.107.124026} {\bibfield  {journal} {\bibinfo  {journal} {\emph {Phys. Rev. D}}\ }\textbf {\bibinfo {volume} {107}},\ \bibinfo {pages} {124026} (\bibinfo {year} {2023})},\ \Eprint {http://arxiv.org/abs/2303.02111} {arXiv:2303.02111 [gr-qc]} \BibitemShut {NoStop}%
\bibitem [{\citenamefont {Olmo}\ \emph {et~al.}(2023)\citenamefont {Olmo}, \citenamefont {Rosa}, \citenamefont {Rubiera-Garcia},\ and\ \citenamefont {Saez-Chillon~Gomez}}]{Olmo:2023lil}%
  \BibitemOpen
  \bibfield  {author} {\bibinfo {author} {\bibfnamefont {G.~J.}\ \bibnamefont {Olmo}}, \bibinfo {author} {\bibfnamefont {J.~L.}\ \bibnamefont {Rosa}}, \bibinfo {author} {\bibfnamefont {D.}~\bibnamefont {Rubiera-Garcia}},  and \bibinfo {author} {\bibfnamefont {D.}~\bibnamefont {Saez-Chillon~Gomez}},\ }\href {\doibase 10.1088/1361-6382/aceacd} {\bibfield  {journal} {\bibinfo  {journal} {\emph {Class. Quant. Grav.}}\ }\textbf {\bibinfo {volume} {40}},\ \bibinfo {pages} {174002} (\bibinfo {year} {2023})},\ \Eprint {http://arxiv.org/abs/2302.12064} {arXiv:2302.12064 [gr-qc]} \BibitemShut {NoStop}%
\bibitem [{\citenamefont {Akiyama}\ \emph {et~al.}(2019)\citenamefont {Akiyama} \emph {et~al.}}]{EventHorizonTelescope:2019dse}%
  \BibitemOpen
  \bibfield  {author} {\bibinfo {author} {\bibfnamefont {K.}~\bibnamefont {Akiyama}} \emph {et~al.} (\bibinfo {collaboration} {Event Horizon Telescope}),\ }\href {\doibase 10.3847/2041-8213/ab0ec7} {\bibfield  {journal} {\bibinfo  {journal} {\emph {Astrophys. J. Lett.}}\ }\textbf {\bibinfo {volume} {875}},\ \bibinfo {pages} {L1} (\bibinfo {year} {2019})},\ \Eprint {http://arxiv.org/abs/1906.11238} {arXiv:1906.11238 [astro-ph.GA]} \BibitemShut {NoStop}%
\bibitem [{\citenamefont {Broderick}\ \emph {et~al.}(2022{\natexlab{a}})\citenamefont {Broderick} \emph {et~al.}}]{Broderick:2022tfu}%
  \BibitemOpen
  \bibfield  {author} {\bibinfo {author} {\bibfnamefont {A.~E.}\ \bibnamefont {Broderick}} \emph {et~al.},\ }\href {\doibase 10.3847/1538-4357/ac7c1d} {\bibfield  {journal} {\bibinfo  {journal} {\emph {Astrophys. J.}}\ }\textbf {\bibinfo {volume} {935}},\ \bibinfo {pages} {61} (\bibinfo {year} {2022}{\natexlab{a}})},\ \Eprint {http://arxiv.org/abs/2208.09004} {arXiv:2208.09004 [astro-ph.HE]} \BibitemShut {NoStop}%
\bibitem [{\citenamefont {Broderick}\ \emph {et~al.}(2022{\natexlab{b}})\citenamefont {Broderick}, \citenamefont {Tiede}, \citenamefont {Pesce},\ and\ \citenamefont {Gold}}]{Broderick:2021ohx}%
  \BibitemOpen
  \bibfield  {author} {\bibinfo {author} {\bibfnamefont {A.~E.}\ \bibnamefont {Broderick}}, \bibinfo {author} {\bibfnamefont {P.}~\bibnamefont {Tiede}}, \bibinfo {author} {\bibfnamefont {D.~W.}\ \bibnamefont {Pesce}},  and \bibinfo {author} {\bibfnamefont {R.}~\bibnamefont {Gold}},\ }\href {\doibase 10.3847/1538-4357/ac4970} {\bibfield  {journal} {\bibinfo  {journal} {\emph {Astrophys. J.}}\ }\textbf {\bibinfo {volume} {927}},\ \bibinfo {pages} {6} (\bibinfo {year} {2022}{\natexlab{b}})},\ \Eprint {http://arxiv.org/abs/2105.09962} {arXiv:2105.09962 [astro-ph.HE]} \BibitemShut {NoStop}%
\bibitem [{\citenamefont {Cunha}\ \emph {et~al.}(2017)\citenamefont {Cunha}, \citenamefont {Berti},\ and\ \citenamefont {Herdeiro}}]{Cunha:2017qtt}%
  \BibitemOpen
  \bibfield  {author} {\bibinfo {author} {\bibfnamefont {P.~V.~P.}\ \bibnamefont {Cunha}}, \bibinfo {author} {\bibfnamefont {E.}~\bibnamefont {Berti}},  and \bibinfo {author} {\bibfnamefont {C.~A.~R.}\ \bibnamefont {Herdeiro}},\ }\href {\doibase 10.1103/PhysRevLett.119.251102} {\bibfield  {journal} {\bibinfo  {journal} {\emph {Phys. Rev. Lett.}}\ }\textbf {\bibinfo {volume} {119}},\ \bibinfo {pages} {251102} (\bibinfo {year} {2017})},\ \Eprint {http://arxiv.org/abs/1708.04211} {arXiv:1708.04211 [gr-qc]} \BibitemShut {NoStop}%
\bibitem [{\citenamefont {Cunha}\ \emph {et~al.}(2023)\citenamefont {Cunha}, \citenamefont {Herdeiro}, \citenamefont {Radu},\ and\ \citenamefont {Sanchis-Gual}}]{Cunha:2022gde}%
  \BibitemOpen
  \bibfield  {author} {\bibinfo {author} {\bibfnamefont {P.~V.~P.}\ \bibnamefont {Cunha}}, \bibinfo {author} {\bibfnamefont {C.}~\bibnamefont {Herdeiro}}, \bibinfo {author} {\bibfnamefont {E.}~\bibnamefont {Radu}},  and \bibinfo {author} {\bibfnamefont {N.}~\bibnamefont {Sanchis-Gual}},\ }\href {\doibase 10.1103/PhysRevLett.130.061401} {\bibfield  {journal} {\bibinfo  {journal} {\emph {Phys. Rev. Lett.}}\ }\textbf {\bibinfo {volume} {130}},\ \bibinfo {pages} {061401} (\bibinfo {year} {2023})},\ \Eprint {http://arxiv.org/abs/2207.13713} {arXiv:2207.13713 [gr-qc]} \BibitemShut {NoStop}%
\bibitem [{\citenamefont {Bozza}(2002)}]{Bozza:2002zj}%
  \BibitemOpen
  \bibfield  {author} {\bibinfo {author} {\bibfnamefont {V.}~\bibnamefont {Bozza}},\ }\href {\doibase 10.1103/PhysRevD.66.103001} {\bibfield  {journal} {\bibinfo  {journal} {\emph {Phys. Rev. D}}\ }\textbf {\bibinfo {volume} {66}},\ \bibinfo {pages} {103001} (\bibinfo {year} {2002})},\ \Eprint {http://arxiv.org/abs/gr-qc/0208075} {arXiv:gr-qc/0208075} \BibitemShut {NoStop}%
\bibitem [{\citenamefont {Keeton}\ and\ \citenamefont {Petters}(2005)}]{Keeton:2005jd}%
  \BibitemOpen
  \bibfield  {author} {\bibinfo {author} {\bibfnamefont {C.~R.}\ \bibnamefont {Keeton}} and \bibinfo {author} {\bibfnamefont {A.~O.}\ \bibnamefont {Petters}},\ }\href {\doibase 10.1103/PhysRevD.72.104006} {\bibfield  {journal} {\bibinfo  {journal} {\emph {Phys. Rev. D}}\ }\textbf {\bibinfo {volume} {72}},\ \bibinfo {pages} {104006} (\bibinfo {year} {2005})},\ \Eprint {http://arxiv.org/abs/gr-qc/0511019} {arXiv:gr-qc/0511019} \BibitemShut {NoStop}%
\bibitem [{\citenamefont {Bozza}(2010)}]{Bozza:2010xqn}%
  \BibitemOpen
  \bibfield  {author} {\bibinfo {author} {\bibfnamefont {V.}~\bibnamefont {Bozza}},\ }\href {\doibase 10.1007/s10714-010-0988-2} {\bibfield  {journal} {\bibinfo  {journal} {\emph {Gen. Rel. Grav.}}\ }\textbf {\bibinfo {volume} {42}},\ \bibinfo {pages} {2269} (\bibinfo {year} {2010})},\ \Eprint {http://arxiv.org/abs/0911.2187} {arXiv:0911.2187 [gr-qc]} \BibitemShut {NoStop}%
\bibitem [{\citenamefont {Cunha}\ and\ \citenamefont {Herdeiro}(2018)}]{Cunha:2018acu}%
  \BibitemOpen
  \bibfield  {author} {\bibinfo {author} {\bibfnamefont {P.~V.~P.}\ \bibnamefont {Cunha}} and \bibinfo {author} {\bibfnamefont {C.~A.~R.}\ \bibnamefont {Herdeiro}},\ }\href {\doibase 10.1007/s10714-018-2361-9} {\bibfield  {journal} {\bibinfo  {journal} {\emph {Gen. Rel. Grav.}}\ }\textbf {\bibinfo {volume} {50}},\ \bibinfo {pages} {42} (\bibinfo {year} {2018})},\ \Eprint {http://arxiv.org/abs/1801.00860} {arXiv:1801.00860 [gr-qc]} \BibitemShut {NoStop}%
\bibitem [{\citenamefont {Shaikh}\ \emph {et~al.}(2019)\citenamefont {Shaikh}, \citenamefont {Banerjee}, \citenamefont {Paul},\ and\ \citenamefont {Sarkar}}]{Shaikh:2019itn}%
  \BibitemOpen
  \bibfield  {author} {\bibinfo {author} {\bibfnamefont {R.}~\bibnamefont {Shaikh}}, \bibinfo {author} {\bibfnamefont {P.}~\bibnamefont {Banerjee}}, \bibinfo {author} {\bibfnamefont {S.}~\bibnamefont {Paul}},  and \bibinfo {author} {\bibfnamefont {T.}~\bibnamefont {Sarkar}},\ }\href {\doibase 10.1103/PhysRevD.99.104040} {\bibfield  {journal} {\bibinfo  {journal} {\emph {Phys. Rev. D}}\ }\textbf {\bibinfo {volume} {99}},\ \bibinfo {pages} {104040} (\bibinfo {year} {2019})},\ \Eprint {http://arxiv.org/abs/1903.08211} {arXiv:1903.08211 [gr-qc]} \BibitemShut {NoStop}%
\bibitem [{\citenamefont {Chowdhuri}\ \emph {et~al.}(2023)\citenamefont {Chowdhuri}, \citenamefont {Ghosh},\ and\ \citenamefont {Bhattacharyya}}]{AbhishekChowdhuri:2023ekr}%
  \BibitemOpen
  \bibfield  {author} {\bibinfo {author} {\bibfnamefont {A.}~\bibnamefont {Chowdhuri}}, \bibinfo {author} {\bibfnamefont {S.}~\bibnamefont {Ghosh}},  and \bibinfo {author} {\bibfnamefont {A.}~\bibnamefont {Bhattacharyya}},\ }\href {\doibase 10.3389/fphy.2023.1113909} {\bibfield  {journal} {\bibinfo  {journal} {\emph {Front. Phys.}}\ }\textbf {\bibinfo {volume} {11}},\ \bibinfo {pages} {1113909} (\bibinfo {year} {2023})},\ \Eprint {http://arxiv.org/abs/2303.02069} {arXiv:2303.02069 [gr-qc]} \BibitemShut {NoStop}%
\bibitem [{\citenamefont {Carter}(1968)}]{Carter:1968rr}%
  \BibitemOpen
  \bibfield  {author} {\bibinfo {author} {\bibfnamefont {B.}~\bibnamefont {Carter}},\ }\href {\doibase 10.1103/PhysRev.174.1559} {\bibfield  {journal} {\bibinfo  {journal} {\emph {Phys. Rev.}}\ }\textbf {\bibinfo {volume} {174}},\ \bibinfo {pages} {1559} (\bibinfo {year} {1968})}\BibitemShut {NoStop}%
\bibitem [{\citenamefont {Chandrasekhar}(1984)}]{Chandrasekhar:1984siy}%
  \BibitemOpen
  \bibfield  {author} {\bibinfo {author} {\bibfnamefont {S.}~\bibnamefont {Chandrasekhar}},\ }\href {\doibase 10.1007/978-94-009-6469-3_2} {\bibfield  {journal} {\bibinfo  {journal} {\emph {Fundam. Theor. Phys.}}\ }\textbf {\bibinfo {volume} {9}},\ \bibinfo {pages} {5} (\bibinfo {year} {1984})}\BibitemShut {NoStop}%
\bibitem [{\citenamefont {Berti}\ \emph {et~al.}(2009)\citenamefont {Berti}, \citenamefont {Cardoso},\ and\ \citenamefont {Starinets}}]{Berti:2009kk}%
  \BibitemOpen
  \bibfield  {author} {\bibinfo {author} {\bibfnamefont {E.}~\bibnamefont {Berti}}, \bibinfo {author} {\bibfnamefont {V.}~\bibnamefont {Cardoso}},  and \bibinfo {author} {\bibfnamefont {A.~O.}\ \bibnamefont {Starinets}},\ }\href {\doibase 10.1088/0264-9381/26/16/163001} {\bibfield  {journal} {\bibinfo  {journal} {\emph {Class. Quant. Grav.}}\ }\textbf {\bibinfo {volume} {26}},\ \bibinfo {pages} {163001} (\bibinfo {year} {2009})},\ \Eprint {http://arxiv.org/abs/0905.2975} {arXiv:0905.2975 [gr-qc]} \BibitemShut {NoStop}%
\bibitem [{\citenamefont {Nakarachinda}\ \emph {et~al.}(2025)\citenamefont {Nakarachinda}, \citenamefont {Boonserm}, \citenamefont {De~Felice}, \citenamefont {Tsujikawa},\ and\ \citenamefont {Wongjun}}]{Nakarachinda:2025bvy}%
  \BibitemOpen
  \bibfield  {author} {\bibinfo {author} {\bibfnamefont {R.}~\bibnamefont {Nakarachinda}}, \bibinfo {author} {\bibfnamefont {P.}~\bibnamefont {Boonserm}}, \bibinfo {author} {\bibfnamefont {A.}~\bibnamefont {De~Felice}}, \bibinfo {author} {\bibfnamefont {S.}~\bibnamefont {Tsujikawa}},  and \bibinfo {author} {\bibfnamefont {P.}~\bibnamefont {Wongjun}},\ }\href {\doibase 10.1103/5bcc-gc31} {\bibfield  {journal} {\bibinfo  {journal} {\emph {Phys. Rev. D}}\ }\textbf {\bibinfo {volume} {112}},\ \bibinfo {pages} {064055} (\bibinfo {year} {2025})},\ \Eprint {http://arxiv.org/abs/2506.18241} {arXiv:2506.18241 [gr-qc]} \BibitemShut {NoStop}%
\bibitem [{\citenamefont {Will}(1994)}]{Will:1994fb}%
  \BibitemOpen
  \bibfield  {author} {\bibinfo {author} {\bibfnamefont {C.~M.}\ \bibnamefont {Will}},\ }\href {\doibase 10.1103/PhysRevD.50.6058} {\bibfield  {journal} {\bibinfo  {journal} {\emph {Phys. Rev. D}}\ }\textbf {\bibinfo {volume} {50}},\ \bibinfo {pages} {6058} (\bibinfo {year} {1994})},\ \Eprint {http://arxiv.org/abs/gr-qc/9406022} {arXiv:gr-qc/9406022} \BibitemShut {NoStop}%
\bibitem [{\citenamefont {Alsing}\ \emph {et~al.}(2012)\citenamefont {Alsing}, \citenamefont {Berti}, \citenamefont {Will},\ and\ \citenamefont {Zaglauer}}]{Alsing:2011er}%
  \BibitemOpen
  \bibfield  {author} {\bibinfo {author} {\bibfnamefont {J.}~\bibnamefont {Alsing}}, \bibinfo {author} {\bibfnamefont {E.}~\bibnamefont {Berti}}, \bibinfo {author} {\bibfnamefont {C.~M.}\ \bibnamefont {Will}},  and \bibinfo {author} {\bibfnamefont {H.}~\bibnamefont {Zaglauer}},\ }\href {\doibase 10.1103/PhysRevD.85.064041} {\bibfield  {journal} {\bibinfo  {journal} {\emph {Phys. Rev. D}}\ }\textbf {\bibinfo {volume} {85}},\ \bibinfo {pages} {064041} (\bibinfo {year} {2012})},\ \Eprint {http://arxiv.org/abs/1112.4903} {arXiv:1112.4903 [gr-qc]} \BibitemShut {NoStop}%
\bibitem [{\citenamefont {Liu}\ \emph {et~al.}(2020)\citenamefont {Liu}, \citenamefont {Zhao},\ and\ \citenamefont {Wang}}]{Liu:2020moh}%
  \BibitemOpen
  \bibfield  {author} {\bibinfo {author} {\bibfnamefont {T.}~\bibnamefont {Liu}}, \bibinfo {author} {\bibfnamefont {W.}~\bibnamefont {Zhao}},  and \bibinfo {author} {\bibfnamefont {Y.}~\bibnamefont {Wang}},\ }\href {\doibase 10.1103/PhysRevD.102.124035} {\bibfield  {journal} {\bibinfo  {journal} {\emph {Phys. Rev. D}}\ }\textbf {\bibinfo {volume} {102}},\ \bibinfo {pages} {124035} (\bibinfo {year} {2020})},\ \Eprint {http://arxiv.org/abs/2007.10068} {arXiv:2007.10068 [gr-qc]} \BibitemShut {NoStop}%
\bibitem [{\citenamefont {Wang}\ \emph {et~al.}(2021)\citenamefont {Wang}, \citenamefont {Li}, \citenamefont {Jiang}, \citenamefont {Yuan}, \citenamefont {Hu},\ and\ \citenamefont {Fan}}]{Wang:2020ori}%
  \BibitemOpen
  \bibfield  {author} {\bibinfo {author} {\bibfnamefont {H.-T.}\ \bibnamefont {Wang}}, \bibinfo {author} {\bibfnamefont {P.-C.}\ \bibnamefont {Li}}, \bibinfo {author} {\bibfnamefont {J.-L.}\ \bibnamefont {Jiang}}, \bibinfo {author} {\bibfnamefont {G.-W.}\ \bibnamefont {Yuan}}, \bibinfo {author} {\bibfnamefont {Y.-M.}\ \bibnamefont {Hu}},  and \bibinfo {author} {\bibfnamefont {Y.-Z.}\ \bibnamefont {Fan}},\ }\href {\doibase 10.1140/epjc/s10052-021-09555-1} {\bibfield  {journal} {\bibinfo  {journal} {\emph {Eur. Phys. J. C}}\ }\textbf {\bibinfo {volume} {81}},\ \bibinfo {pages} {769} (\bibinfo {year} {2021})},\ \Eprint {http://arxiv.org/abs/2004.12421} {arXiv:2004.12421 [gr-qc]} \BibitemShut {NoStop}%
\bibitem [{\citenamefont {Niu}\ \emph {et~al.}(2021)\citenamefont {Niu}, \citenamefont {Zhang}, \citenamefont {Wang},\ and\ \citenamefont {Zhao}}]{Niu:2021nic}%
  \BibitemOpen
  \bibfield  {author} {\bibinfo {author} {\bibfnamefont {R.}~\bibnamefont {Niu}}, \bibinfo {author} {\bibfnamefont {X.}~\bibnamefont {Zhang}}, \bibinfo {author} {\bibfnamefont {B.}~\bibnamefont {Wang}},  and \bibinfo {author} {\bibfnamefont {W.}~\bibnamefont {Zhao}},\ }\href {\doibase 10.3847/1538-4357/ac1d4f} {\bibfield  {journal} {\bibinfo  {journal} {\emph {Astrophys. J.}}\ }\textbf {\bibinfo {volume} {921}},\ \bibinfo {pages} {149} (\bibinfo {year} {2021})},\ \Eprint {http://arxiv.org/abs/2105.13644} {arXiv:2105.13644 [gr-qc]} \BibitemShut {NoStop}%
\bibitem [{\citenamefont {Higashino}\ and\ \citenamefont {Tsujikawa}(2023)}]{Higashino:2022izi}%
  \BibitemOpen
  \bibfield  {author} {\bibinfo {author} {\bibfnamefont {Y.}~\bibnamefont {Higashino}} and \bibinfo {author} {\bibfnamefont {S.}~\bibnamefont {Tsujikawa}},\ }\href {\doibase 10.1103/PhysRevD.107.044003} {\bibfield  {journal} {\bibinfo  {journal} {\emph {Phys. Rev. D}}\ }\textbf {\bibinfo {volume} {107}},\ \bibinfo {pages} {044003} (\bibinfo {year} {2023})},\ \Eprint {http://arxiv.org/abs/2209.13749} {arXiv:2209.13749 [gr-qc]} \BibitemShut {NoStop}%
\bibitem [{\citenamefont {Takeda}\ \emph {et~al.}(2024)\citenamefont {Takeda}, \citenamefont {Tsujikawa},\ and\ \citenamefont {Nishizawa}}]{Takeda:2023wqn}%
  \BibitemOpen
  \bibfield  {author} {\bibinfo {author} {\bibfnamefont {H.}~\bibnamefont {Takeda}}, \bibinfo {author} {\bibfnamefont {S.}~\bibnamefont {Tsujikawa}},  and \bibinfo {author} {\bibfnamefont {A.}~\bibnamefont {Nishizawa}},\ }\href {\doibase 10.1103/PhysRevD.109.104072} {\bibfield  {journal} {\bibinfo  {journal} {\emph {Phys. Rev. D}}\ }\textbf {\bibinfo {volume} {109}},\ \bibinfo {pages} {104072} (\bibinfo {year} {2024})},\ \Eprint {http://arxiv.org/abs/2311.09281} {arXiv:2311.09281 [gr-qc]} \BibitemShut {NoStop}%
\bibitem [{\citenamefont {Hinderer}(2008)}]{Hinderer:2007mb}%
  \BibitemOpen
  \bibfield  {author} {\bibinfo {author} {\bibfnamefont {T.}~\bibnamefont {Hinderer}},\ }\href {\doibase 10.1086/533487} {\bibfield  {journal} {\bibinfo  {journal} {\emph {Astrophys. J.}}\ }\textbf {\bibinfo {volume} {677}},\ \bibinfo {pages} {1216} (\bibinfo {year} {2008})},\ \bibinfo {note} {[Erratum: Astrophys.J. 697, 964 (2009)]},\ \Eprint {http://arxiv.org/abs/0711.2420} {arXiv:0711.2420 [astro-ph]} \BibitemShut {NoStop}%
\bibitem [{\citenamefont {De}\ \emph {et~al.}(2018)\citenamefont {De}, \citenamefont {Finstad}, \citenamefont {Lattimer}, \citenamefont {Brown}, \citenamefont {Berger},\ and\ \citenamefont {Biwer}}]{De:2018uhw}%
  \BibitemOpen
  \bibfield  {author} {\bibinfo {author} {\bibfnamefont {S.}~\bibnamefont {De}}, \bibinfo {author} {\bibfnamefont {D.}~\bibnamefont {Finstad}}, \bibinfo {author} {\bibfnamefont {J.~M.}\ \bibnamefont {Lattimer}}, \bibinfo {author} {\bibfnamefont {D.~A.}\ \bibnamefont {Brown}}, \bibinfo {author} {\bibfnamefont {E.}~\bibnamefont {Berger}},  and \bibinfo {author} {\bibfnamefont {C.~M.}\ \bibnamefont {Biwer}},\ }\href {\doibase 10.1103/PhysRevLett.121.091102} {\bibfield  {journal} {\bibinfo  {journal} {\emph {Phys. Rev. Lett.}}\ }\textbf {\bibinfo {volume} {121}},\ \bibinfo {pages} {091102} (\bibinfo {year} {2018})},\ \bibinfo {note} {[Erratum: Phys.Rev.Lett. 121, 259902 (2018)]},\ \Eprint {http://arxiv.org/abs/1804.08583} {arXiv:1804.08583 [astro-ph.HE]} \BibitemShut {NoStop}%
\bibitem [{\citenamefont {Diedrichs}\ \emph {et~al.}(2025)\citenamefont {Diedrichs}, \citenamefont {Tsujikawa},\ and\ \citenamefont {Yagi}}]{Diedrichs:2025vhv}%
  \BibitemOpen
  \bibfield  {author} {\bibinfo {author} {\bibfnamefont {R.~F.}\ \bibnamefont {Diedrichs}}, \bibinfo {author} {\bibfnamefont {S.}~\bibnamefont {Tsujikawa}},  and \bibinfo {author} {\bibfnamefont {K.}~\bibnamefont {Yagi}},\ }\href {\doibase 10.1103/cmb4-chn3} {\bibfield  {journal} {\bibinfo  {journal} {\emph {Phys. Rev. D}}\ }\textbf {\bibinfo {volume} {112}},\ \bibinfo {pages} {044023} (\bibinfo {year} {2025})},\ \Eprint {http://arxiv.org/abs/2501.07998} {arXiv:2501.07998 [gr-qc]} \BibitemShut {NoStop}%
\end{thebibliography}%

\end{document}